\documentclass[reqno, 11pt]{amsart}
\usepackage[utf8]{inputenc}
\usepackage{amsmath,amssymb}
\usepackage[sort&compress]{natbib}
\usepackage[version=3]{mhchem}
\usepackage{tikz}
\usepackage{footmisc}
\usepackage{enumerate}
\usepackage{mathrsfs}
\usepackage{booktabs}
\usepackage{mathtools}
\usepackage{graphicx}
\usepackage{dcolumn}
\usepackage{bm}
\usepackage{algorithm}
\usepackage{algpseudocode}
\usepackage{amsaddr}
\usepackage[margin=1.3in]{geometry}

\renewcommand{\phi}{\varphi}
\renewcommand{\theta}{\vartheta}

\renewcommand{\b}{\mathbf}
\newcommand{\adj}{^\dagger}

\newcommand{\ket}[1]{\vert #1 \rangle}
\newcommand{\Ket}[1]{\Big\vert #1 \Big\rangle}
\newcommand{\bra}[1]{\langle #1 \vert}
\newcommand{\Bra}[1]{\Big\langle #1 \Big\vert}

\newcommand{\Tbraket}[3]{\langle #1 \hspace{.10em} \vert \hspace{.10em} #2 \hspace{.10em} \vert \hspace{.10em} #3 \rangle}

\newcommand{\ident}{\mathbb{I}}

\newcommand{\cre}[1]{a_{#1}\adj}
\newcommand{\ann}[1]{a_{#1}}

\newcommand{\sg}[2]{\begin{array}{c}
#2 \\ #1
\end{array}}
\newcommand{\db}[4]{\begin{array}{c}
#2#4 \\ #1#3
\end{array}}

\newcommand{\J}{\b{A}}

\newcommand{\nucl}{\b{R}}

\newcommand{\veta}{\boldsymbol{\eta}} 

\newcommand{\vq}{\b{q}}

\newcommand{\bsym}{\boldsymbol} 

\begin{document}
\title[Conical intersections for coupled cluster dynamics]{Resolving the notorious case of conical intersections for coupled cluster dynamics}
\author{Eirik F.~Kj{\o}nstad and Henrik Koch$^\ast$}\footnote{henrik.koch@ntnu.no}
\address{Department of Chemistry, Norwegian University of Science and Technology, 7491 Trondheim, Norway} 
\address{Department of Chemistry and the PULSE Institute, Stanford University, Stanford, California 94305, USA}
\begin{abstract} The motion of electrons and nuclei in photochemical events often involve conical intersections, degeneracies between electronic states. They serve as funnels for nuclear relaxation---on the femtosecond scale---in processes where the electrons and nuclei couple nonadiabatically. Accurate \emph{ab initio} quantum chemical models are essential for interpreting experimental measurements of such phenomena. In this paper we resolve a long-standing problem in  coupled cluster theory, presenting the first formulation of the theory that correctly describes conical intersections between excited electronic states of the same symmetry. This new development demonstrates that the highly accurate coupled cluster theory can be applied to describe dynamics on excited electronic states involving conical intersections.\end{abstract}
\maketitle

Conical intersections, or electronic degeneracies, are widely recognized as central to the motion of nuclei and electrons in photochemical events\citep{Matsika2011,Zhu2016}. They have been implicated in a range of chemical reactions, from the ring-opening reaction of 1,3-cyclohexadiene\citep{Kim2015} and the proton transfer reaction in hydroxybenzaldehyde\citep{Migani2008} to the cis-trans isomerization thought to be the primary photochemical event in human vision\citep{Polli2010}. Our understanding of nuclear dynamics is firmly rooted in the often accurate Born-Oppenheimer approximation, where the motion of the electrons create potential energy surfaces to which the nuclei in turn respond\citep{Born1927}. However, the approximation breaks down completely when a molecule approaches a conical intersection, giving rise to femtosecond ($10^{-15}$ s) processes that involve an intricate interplay between nuclear and electronic motion\citep{Ben2002}. Pump-probe techniques have made these ultrafast processes increasingly amenable to experimental investigation\citep{Stolow2003}.

Rapid developments in \emph{ab initio} quantum chemistry was spurred by the realization that nonadiabaticity is the norm in photochemistry. These include assessments of the potential energy surfaces close to electronic degeneracies\citep{Gozem2014,Tuna2015}, attempts to incorporate nonadiabaticity in dynamics simulations by solving the time-dependent Schrödinger equation explicitly\citep{Ben2000,Ben2002}, and implementations of the nonadibatic coupling elements predicted by various \emph{ab initio} models\citep{Bak1992,Lischka2004,Tajti2009}. A major overarching goal of this research is the reliable prediction of nonadiabatic dynamics, which will enable one to monitor, in real-time, processes---such as electron density fluctuations\citep{Ben2002}---not directly accessible by experiment\citep{Matsika2011}. Since the Schrödinger equation cannot be solved exactly for many-electron molecular systems, such predictions must neccessarily be grounded in approximate treatments of electronic correlation, or electron-electron interactions\citep{Tew2007}.

The most successful treatment of electronic correlation is provided by coupled cluster theory\citep{Coester1958,Cizek1966}, a model now routinely applied to chemically interesting systems in spite of its steep computational scaling\citep{Bartlett2007}. Nevertheless, as is true for all quantum chemical models, the theory is not globally accurate and may fail to describe certain regions of the potential energy surfaces. For instance, the accuracy of its ground state wavefunction diminishes in regions of nuclear space where the exact wavefunction has multireference character (\emph{e.g.}, when a molecule dissociates into fragments)\citep{Lyakh2012}. In the following, we restrict ourselves to the excited electronic states, where multireference character is not an issue\citep{Krylov2008}, but where other problems have hindered its successful application to conical intersections.   
About a decade ago, Hättig\citep{Hattig2005} argued that matrix symmetry---a property that coupled cluster theory does not have---is necessary to properly describe conical intersections between states of the same symmetry. 
The unphysical complex energies predicted to exist due to nonsymmetry\citep{Hattig2005} were later reported in coupled cluster singles and doubles (CCSD)\citep{Purvis1982} and triples (CCSDT)\citep{Noga1987,Noga1988} calculations\citep{Kohn2007}. More recently, complex energies were encountered in dynamics simulations using the perturbative doubles (CC2)\citep{Christiansen1995} model, illustrating the relevance of these issues in realistic applications\citep{Plasser2014}. Moreover, since same-symmetry conical intersections represent the vast majority of degeneracies\citep{Zhu2016}, coupled cluster theory has been of limited use for conical intersections in general.

The state-of-the-art theories for conical intersections are complete active space (CAS) models, such as CASSCF and CASPT2\citep{Andersson1990}. Although they give a physically correct description of conical intersections by their matrix symmetry (see, \emph{e.g.}, Ben-Nun \emph{et al.}\citep{Ben2000}), their ability to account for dynamic correlation is limited\citep{Levine2008}. The same can be said for the algebraic diagrammatic construction (ADC)\citep{Schirmer1982} theory advocated by some groups\citep{Hattig2005,Dreuw2015}. The ground state wavefunction in ADC, obtained by Møller-Plesset perturbation theory, is known to have a limited domain of validity\citep{Christiansen1996,Helgaker2014}. For large systems, computational chemists often resort to density functional theory (DFT), which is less computationally demanding than \emph{ab initio} theories, but also less accurate\citep{Levine2006,Picconi2011}. On the other hand, coupled cluster theory accurately accounts for dynamic correlation effects and multireference character in excited states. A formulation of the theory able to treat conical intersections will therefore be a highly desirable addition to current methodologies. 

For instance, theoretical investigations by various methods have provided inconsistent predictions for the $\pi \pi^\ast$ nuclear relaxation in thymine\citep{Improta2016}. Some simulations predict that relaxation proceeds first to a local minimum of the $\pi \pi^\ast$ state within 100 femtoseconds, followed by slow internal conversion from the $\pi \pi^\ast$ state to the $n\pi^\ast$ state (CASPT2)\citep{Hudock2007} or to the ground state (CASSCF)\citep{Asturiol2009}; others predict fast barrierless $\pi\pi^\ast/n\pi^\ast$ relaxation (TD-DFT)\citep{Picconi2011} or direct $\pi\pi\ast$/ground state relaxation within a few hundred femtoseconds (CASPT2)\citep{Nakayama2013}. Evidently, accurate quantum chemical predictions are essential for reliable predictions in dynamics simulations. The accuracy of coupled cluster theory was recently shown in experiments confirming ultrafast $\pi\pi^\ast/n\pi^\ast$ conversion\citep{Wolf2017}. This study emphasizes the need for highly accurate methods in excited state dynamics.

In a recent paper, we showed that nonsymmetric theories provide a correct description of conical intersections if they are nondefective, a mathematical property that ensures nonparallel eigenstates\citep{Kjoenstad2017b}. Here we demonstrate that coupled cluster theory can be constrained to be nondefective, thereby resolving the long-standing intersection issues\citep{Hattig2005}. The modified formulation of the theory, named similarity constrained coupled cluster theory, provides a correct description of same-symmetry conical intersections. In particular, we illustrate numerically that this is the case for a conical intersection in hypofluorous acid. This new development shows that coupled cluster theory can be applied to nonadiabatic photochemical processes.

\section*{Theory} 
The coupled cluster ground state wavefunction is written ${\ket{\Psi} = e^T \, \ket{\Phi_0}}$ for the Hartree-Fock state $\ket{\Phi_0}$, where ${T = \sum t_\mu \, \tau_\mu}$, the cluster operator, consists of excitation operators, $\tau_\mu$, weighted by amplitudes, $t_\mu$\citep{Cizek1966,Cizek1971}. The $n$th excitation energy and electronic state, $\omega_n$ and $\b{r}_n$, are determined from
\begin{align}
\J \, \b{r}_n = \omega_n \, \b{r}_n, \quad A_{\mu\nu} = \Tbraket{\Phi_\mu}{(\hat{H}-E_0)}{\Phi_\nu}, \quad \ket{\Phi_\mu} = \tau_\mu \, \ket{\Phi_0}, \quad \tau_0 = \ident,
\end{align}
where $E_0 = \Tbraket{\Phi_0}{\hat{H}}{\Phi_0}$ and $\hat{H} = e^{-T}  H \, e^T$ is the similarity transformed Hamiltonian\citep{Koch1990,Stanton1993a}. In the CCSD model, the cluster operator is restricted to one- and two-electron excitations, with amplitudes determined by projection onto the corresponding excited determinants\citep{Purvis1982}.

\begin{figure}[htb]
\begin{tikzpicture}[scale=1.2]
\draw[thick] (0.5,0) .. controls (0,4) and (4,0) .. (4,4);
\draw[thick] (0.7,0) .. controls (0.2,3.6) and (4.2,-0.2) .. (4.2,4);
\draw[thick,fill=gray!50] (0.6,0) ellipse (0.10cm and 0.08cm);
\draw[thick,fill=gray!50] (4.1,4) ellipse (0.10cm and 0.08cm);
\draw[thick] (5.5,0) .. controls (5,4) and (9,0) .. (9,4);
\draw[thick] (7,2) -- (7.5,3);
\draw[thick] (7,2) -- (6.5,3);
\draw[thick] (7,2) -- (7.5,1);
\draw[thick] (7,2) -- (6.5,1);
\draw[thick,fill=gray!20] (7,3) ellipse (0.5cm and 0.1cm);
\draw[thick,fill=gray!20] (7,1) ellipse (0.5cm and 0.1cm);
\draw[thick] (2,2.9) .. controls (2.5,1.74) .. (3,2.9);
\draw[thick] (2,1.1) .. controls (2.5,1.98) .. (3,1.1);
\draw[thick,fill=gray!20] (2.5,2.9) ellipse (0.5cm and 0.1cm);
\draw[thick,fill=gray!20] (2.5,1.1) ellipse (0.5cm and 0.1cm);
\draw[thin,->] (0,3) .. controls (1,2) .. (1.7,1.85);
\begin{scope}[scale=0.6,transform shape]
\node at (0.45,5.95) {Complex pair of  energies};
\node at (0.35,5.5) {in the interior of the cylinder};
\end{scope}
\begin{scope}[scale=0.6,transform shape]
\node at (17,6.95) {Intersection seam};
\node at (17,6.5) {of dimension $N-2$};
\end{scope}
\begin{scope}[scale=0.6,transform shape]
\node at (9.2,6.95) {Intersection cylinder};
\node at (9.2,6.5) {of dimension $N-1$};
\end{scope}
\begin{scope}[scale=0.6,transform shape]
\node at (3.7,0.45) {Apparent intersection seam};
\node at (3.7,0) {of dimension $N-2$};
\end{scope}
\begin{scope}[scale=0.7,transform shape]
\node at (3,-1.4) {\textbf{Unconstrained}};
\node at (3,-1.9) {\textbf{coupled cluster theory}};
\node at (10.5,-1.4) {\textbf{Constrained}};
\node at (10.5,-1.9) {\textbf{coupled cluster theory}};
\end{scope}
\end{tikzpicture}
\caption{\label{fig:cc_intersections} \footnotesize Conical intersections in coupled cluster theory. The illustrated shapes are for three vibrational degrees of freedom ($N = 3$), appropriate for hypofluorous acid and other three-atomic molecules. Superimposed on the illustrated vibrational space are potential energy surfaces in the plane orthogonal to the seams at a point of intersection.}
\end{figure}
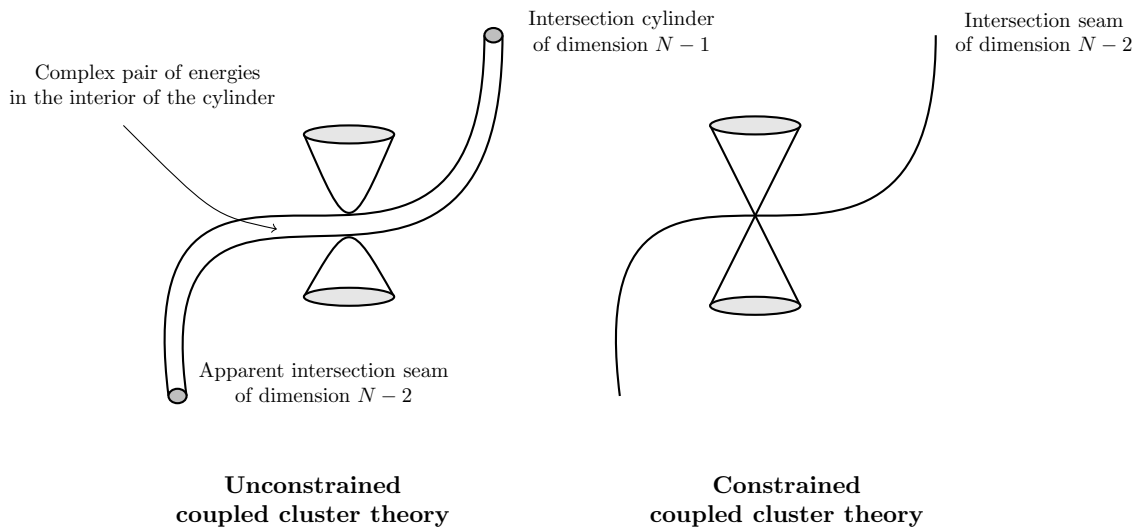

\subsection*{The origin of intersection artifacts} \label{sec:artifacts} We recently traced the unphysical artifacts, observed using coupled cluster methods at same-symmetry conical intersections\citep{Kohn2007}, to defects in the nonsymmetric matrix $\J$\citep{Kjoenstad2017b}. Matrices are known as defective when they are impossible to diagonalize, \emph{i.e.}~when two or more of their eigenvectors are parallel\citep{Golub2012}. 

Considering a representation of $\J$ in a basis of the intersecting states, $J_{ij} (\nucl)$, $i, j = 1,2$, where $\nucl$ is a nuclear coordinate,
Hättig\citep{Hattig2005} argued that at a degeneracy of a nondefective and nonsymmetric $\b{J}$,
\begin{align}
J_{11} = J_{22}, \quad J_{12} = 0, \quad J_{21} = 0, \label{eq:three_conditions}
\end{align}
and concluded, as others have since\citep{Kohn2007,Gozem2014}, that the intersections of coupled cluster theory are qualitatively wrong. This is because equation \eqref{eq:three_conditions} has one more condition than in quantum mechanics\citep{Teller1937,vonNeumann1929}, and may be expected to give intersections of the dimension $N-3$, where $N$ is the number of vibrational degrees of freedom. These three conditions are redundant for nondefective matrices, however. It can be shown that the $\nucl$ satisfying them are expected to inhabit a space of the correct dimension\citep{Kjoenstad2017b} $N-2$.

In practice, $\J$ is defective with intersections where ${(J_{11}-J_{22})^2 + 4 \, J_{12} \,J_{21} = 0}$, an equation obeyed in a space of dimension\citep{Hattig2005} ${N-1}$. While this dimensionality is incorrect, the degeneracy is folded on itself. The intersection is a cylinder instead of a curve for ${N = 3}$, for instance, resembling a seam of dimension $N-2$. See Figure \ref{fig:cc_intersections}. Inside the cylinder the excitation energies are complex and on its surface $\J$ is defective\citep{Kohn2007,Kjoenstad2017b}.

\subsection*{Similarity constrained coupled cluster theory}

When the cluster operator is complete (\emph{i.e.}~includes every excitation), $\J + E_0 \, \b{I}$ is mathematically similar to a representation, $\b{H}$, of the clamped-nuclei Hamiltonian $H$. It can be shown that if $\b{c}_n$ is an eigenvector of $\b{H}$, then $\b{c}_n = \b{Q} \, \b{r}_n$, where $Q_{\mu\nu} = \Tbraket{\Phi_\mu}{e^T}{\Phi_\nu}$. The orthogonality of the $\b{c}_n$, implied by the Hermiticity of $\b{H}$, translates to a generalized orthogonality for the $\b{r}_n$:
\begin{align}
\b{c}_k^\mathrm{T} \b{c}_l = \b{r}_k^\mathrm{T} \b{Q}^\mathrm{T}  \b{Q} \, \b{r}_l = 0, \quad k \neq l. \label{eq:generalized_orthonormality}
\end{align}
As this relation is only satisfied for a complete cluster operator, some of the eigenvectors $\{\b{r}_k\}_k$ may and indeed do become parallel at same-symmetry intersections in truncated coupled cluster methods. In the full space limit, the left eigenvectors $\b{l}_k$ and $\b{l}_l$ are similarly orthogonal over the inverse, $(\b{Q}^\mathrm{T}  \b{Q})^{-1}$, of the above metric.

The wavefunction of similarity constrained CCSD (SCCSD) is defined by including an additional triple excitation in cluster operator $\mathcal{T}$:
\begin{align}
\mathcal{T} = \sum_{ai} t_i^a \, \tau_i^a + \frac{1}{2}\sum_{aibj} t_{ij}^{ab} \, \tau_{ij}^{ab} + \zeta \, \tau_{IJK}^{ABC}.
\end{align}
The amplitudes $t_i^a$, $t_{ij}^{ab}$, and $\zeta$ are determined such that $i$) equation ~\eqref{eq:generalized_orthonormality} is valid for the two intersecting states, and $ii$) the projected equations of the CCSD model are satisfied. This leads to a coupled set of equations that may be solved self-consistently.

Since $\b{Q}^\mathrm{T}  \b{Q}$ is positive definite, parallel eigenvectors cannot satisfy equation \eqref{eq:generalized_orthonormality}, and thus the theory is nondefective. For detailed descriptions of the implementation, we refer to the Supporting Information. 

\begin{figure}[htb]
\includegraphics[width=0.95\linewidth]{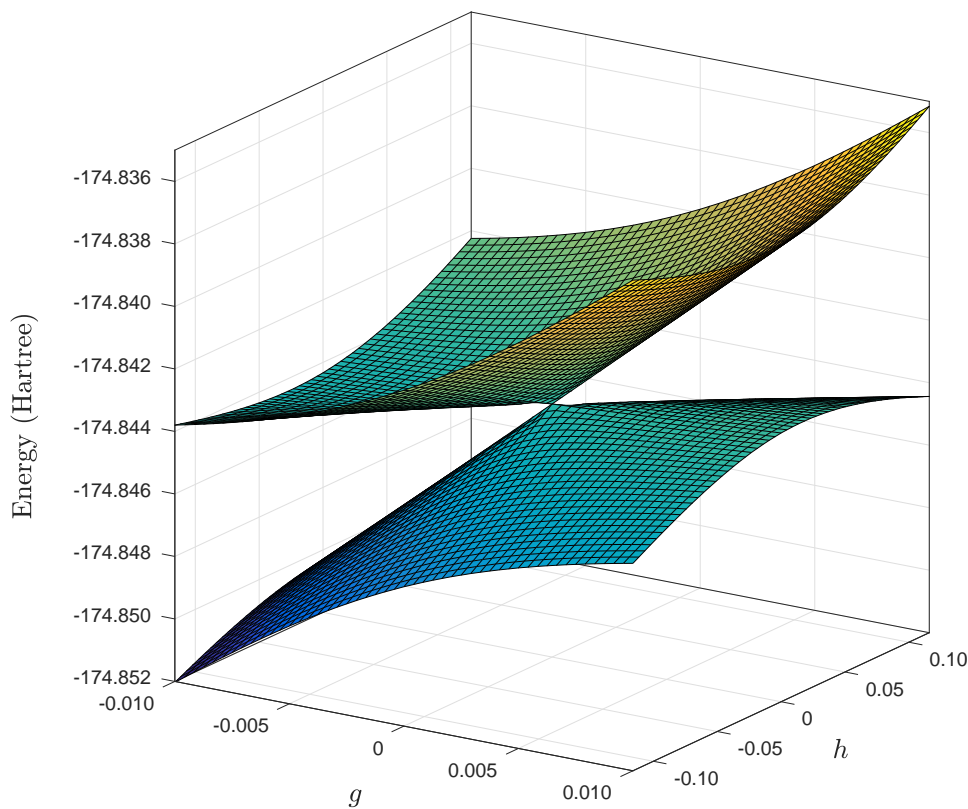}
\caption{\footnotesize A plot of a branching plane in hypofluorous acid (SCCSD/aug-cc-pVDZ) between the $2 \, {^{1}}A'$ and $3 \, {^{1}}A'$  excited states.} \label{fig:hof}
\end{figure}

\section*{Results and Discussion} 

\subsection*{The $2 \, {^{1}}A_1 / 3 \, {^{1}}A_1$ intersection in formaldehyde} 
\begin{figure}[htb]
\includegraphics[width=0.9\linewidth]{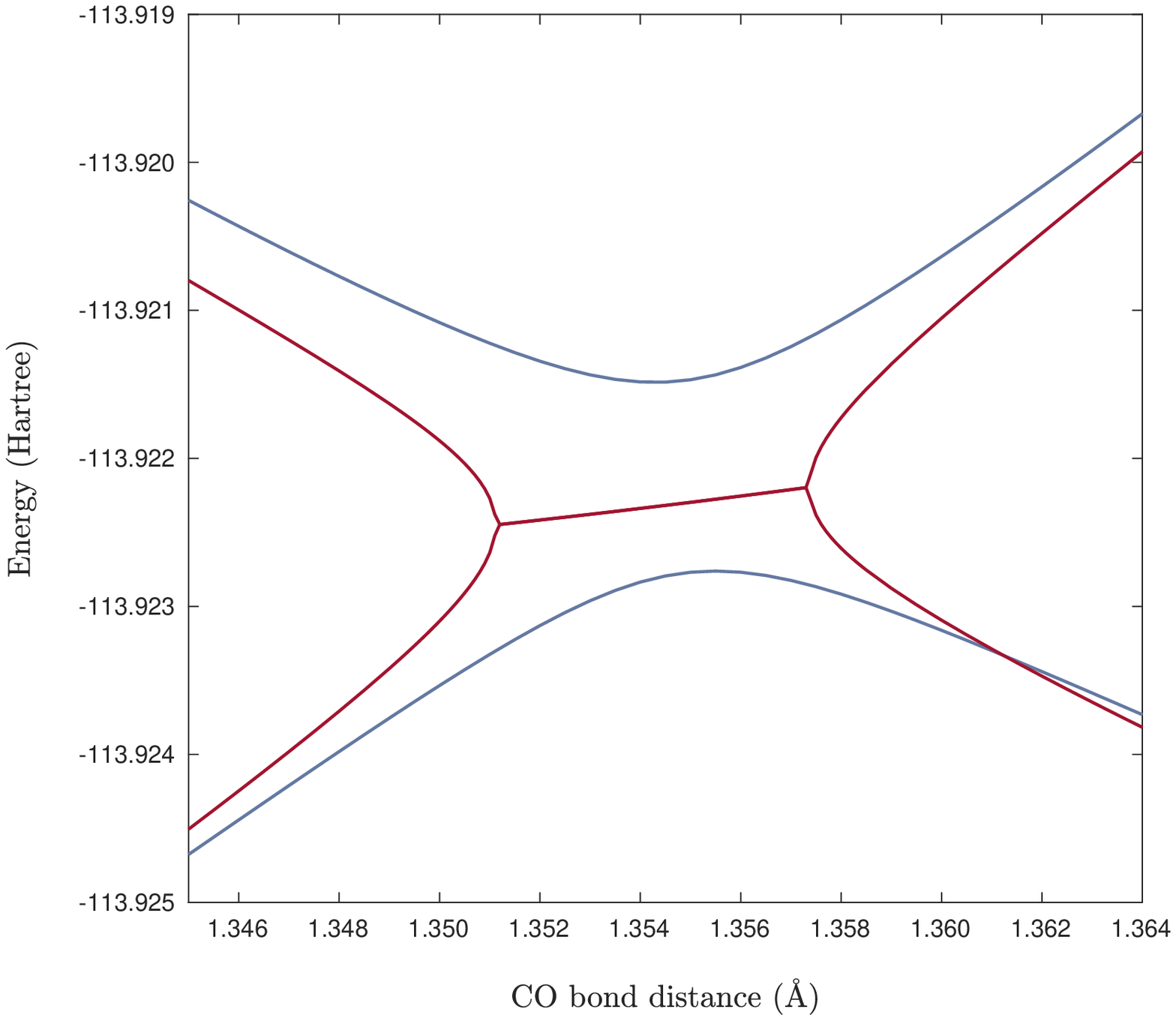}
\caption{\footnotesize The $2 \, {^{1}}A_1$ and $3 \, {^{1}}A_1$ excited states of formaldehyde using CCSD (red) and SCCSD (blue) with an aug-cc-pVDZ basis. The real part of the CCSD energies is shown. An imaginary pair of energies is obtained for the \ce{C=O} bond distances 1.3515--1.3570 Å.} \label{fig:formaldehyde}
\end{figure}

As first shown by Köhn and Tajti\citep{Kohn2007}, the lowest singlet excited states of $A_1$ symmetry in formaldehyde have a defective conical intersection. Here we reproduce their findings, and compare them with the predictions of the similarity constrained theory. The results are shown in Figure \ref{fig:formaldehyde}, where we have used the same geometry as in the original study\citep{Kohn2007}.

The unphysical behavior of CCSD is evident. The states become degenerate at $1.3515 \; \text{\AA}$ and $1.3570\; \text{\AA}$,  giving a complex pair of energies in-between ($E_\pm = E_\mathrm{real} \pm i \, E_\mathrm{imag}$). Such artifacts are absent in the constrained model, where the defective intersection becomes an avoided crossing. If it exists, the conical intersection of the theory is located elsewhere in nuclear space.

\subsection*{The $2 \, {^{1}}A' / 3 \, {^{1}}A'$ intersection in hypofluorous acid} The lowest singlet excited states of $A'$ symmetry in hypoclorous acid intersect\citep{Nanbu1992}. We investigate this conical intersection in the related hypofluorous acid. This three-atomic molecule (\ce{H-O-F}) provides a pedagogical illustration of the model, allowing direct comparisons with Figure \ref{fig:cc_intersections}. After locating a point of intersection, we performed a scan in the branching plane, the plane orthogonal to the intersection seam\citep{Yarkony2001b}. The results are given in Figure \ref{fig:hof}. 

In a recent paper, we showed that nondefective coupled cluster models exhibit the first-order branching plane energy gap linearity\citep{Kjoenstad2017b}, that is, they are conical\citep{Yarkony2001b}. From Figure \ref{fig:hof}, we see that energy gap linearity is obeyed, as the energy surfaces exhibit the physically correct conical appearance. Note the contrast to calculations on formaldehyde using CCSD, where the energy gradient changes more rapidly close to the intersection; this can be seen from Figure \ref{fig:formaldehyde} in the vicinity of the CO bond distances $1.3515 \; \text{\AA}$ and $1.3570 \; \text{\AA}$.

\subsection*{The cluster operator} 
In this initial development, we have used a conceptually simple and size-intensive\citep{Koch1990b} cluster operator $\mathcal{T}$, adding only one additional triple excitation. As we are aiming for a small correction of the wavefunction $\ket{\Psi} = e^\mathcal{T} \, \ket{\Phi_0}$, the excitation is selected, from the dominant single and double excitations contributing to the two states, such that the $\zeta$ parameter is sufficiently small (we used $2$ as a threshold). The excitation is selected at a particular geometry and kept unchanged in subsequent calculations. This selection procedure can easily be made black-box; in the initial geometry of the simulation, an appropriate excitation is identified from an automated test of several excitations. 

We have considered variations in the energies for twelve choices of triple excitation, see Table 1 in the Supporting Information. The energies are found to differ from CCSD by less than five milliHartrees for all excitations, and the energy gaps are similar to the CCSD and CC3 gaps (but different from the CC2 gap). In terms of quality and the location of the intersection seam, the constrained model is thus similar but not identical to CCSD. This behavior is as expected this close to seam, given the proximity to the unphysical cylinder (see Figure \ref{fig:cc_intersections}). A detailed numerical illustration of the cylinder is reported in our recent paper\citep{Kjoenstad2017b}. 

Other formulations are indeed possible where the cluster operator is both orbital and state invariant. For instance, orbital invariance is obtained if the triples contribution in $\mathcal{T}$ is the product of the singles and doubles contributions in one of the states. This also guarantees continuity of the potential energy surfaces. Let $r_{\mu_1}$ and $r_{\mu_2}$ denote the singles and doubles contributions in the state $\b{r}$. The cluster operator is then defined as
\begin{align}
\mathcal{T} = T_1 + T_2 + \zeta \, \sum_{\mu_1 \mu_2} r_{\mu_1} r_{\mu_2} \tau_{\mu_1}\tau_{\mu_2}, 
\end{align}
where the $\zeta$ parameter is used to enforce orthogonality. In this equation, only one of the intersecting states is selected. However, both can be included to give a balanced, or state invariant, operator. Alternatively, the states can provide two parameters, $\zeta_1$ and $\zeta_2$, which will allow generalized orthogonality to be enforced between the left and right eigenvectors simultaneously. For the theory to be nondefective, however, orthogonality between either sets of eigenvectors is sufficient.

\subsection*{Future outlook and conclusions} Similarity constrained coupled cluster theory (SCCSD) gives a physically correct description of an $2 \, {^{1}}A' / 3 \, {^{1}}A'$ same-symmetry conical intersection in hypofluorous acid, with both the proper dimensionality of the intersection seam as well as the correct energy gap linearity in the branching plane\citep{Zhu2016}. Confirming our predictions from a recent paper\citep{Kjoenstad2017b} and resolving a long-standing problem in coupled cluster theory\citep{Hattig2005}, this finding demonstrates that the model can properly describe, with minor modifications, same-symmetry conical intersections between electronic excited states.

Nonadiabatic coupling elements and energy gradients govern the dynamics close to a conical intersection\citep{Worth2004}. Implementing these quantities in the similarity constrained theory is thus necessary for it to be applied in \emph{ab initio} dynamics simulations. These developments are within reach in the near future, although some controversies for the nonadiabatic coupling elements remains to be settled\citep{Christiansen1999,Tajti2009}. On the other hand, techniques for energy gradients are well-established\citep{Koch1990b,Stanton1993b}. For use in dynamics simulations on larger systems, the model should be extended to the lower levels in the coupled cluster hierarchy. Particularly relevant is a perturbative doubles model (SCC2, analogous to CC2\citep{Christiansen1995}), which should scale as $N^5$, where $N$ is the number of orbitals. As the approach is further developed, we will also implement the cluster operators which ensure state and orbital invariance.

As a closing remark, we note that the approach adopted in this paper, namely to enforce a feature of the exact wavefunction (\emph{i.e.}, nondefectiveness), could potentially have more wide-reaching applications. The standard philosophy in \emph{ab initio} quantum chemistry is to solve ever more accurate representations of the Schrödinger equation (the coupled cluster hierarchy results by expanding the subspace onto which the equation is projected). Yet many desirable features of the wavefunction---such as gauge invariance, the related origin invariance\citep{Pedersen1997}, and the correct scaling of molecular properties and transition moments\citep{Koch1990b,Kobayashi1994}---are only valid for a complete cluster operator, and in some cases a complete one-electron basis. Constraining the approximate wavefunction to satisfy exact properties by adding extra correlation (\emph{e.g.}, triple excitation operators in a CCSD formalism) may turn out to be very useful.

\section*{Methods} 
 \footnotesize Calculations were carried out using the Dalton quantum chemistry program\citep{DALTON}. We converged energies and residuals to within $10^{-8}$. The orthogonality in equation \eqref{eq:generalized_orthonormality} was converged to within $10^{-6}$, giving energies correct to approximately $10^{-6}$ Hartrees. The energies in Figures \ref{fig:hof} and \ref{fig:formaldehyde} are obtained by ${\tau_{IJK}^{ABC} = \tau_{7,5,8}^{10,2,2}}$ and ${\tau_{IJK}^{ABC} = \tau_{8,8,7}^{7,2,3}}$, respectively, where the canonical orbitals are ordered according to their energy, from low to high. Complex energies were converged with a modified Davidson algorithm\citep{Kohn2007}. 

A branching plane in hypofluorous acid was identified as follows. First we located a point of intersection. By searching along three orthonormal vibrational coordinates, we then found the intersection seam vector $\b{s}$, the direction in which the degeneracy is preserved. To obtain an orthogonal basis of the branching plane (the orthogonal complement to $\b{s}$), we chose the direction in which the energy difference increased most (denoted $\b{g}$, where $\b{g} \perp \b{s}$) and the vector orthogonal to $\b{s}$ and $\b{g}$ (denoted $\b{h}$). The normal modes used in the above procedure, as well as cartesian coordinates of $\b{s}$, $\b{g}$, $\b{h}$, and of the intersection geometry $\b{R}_0$, are given in the Supporting Information.

For formaldehyde, we performed the scan $R_{\ce{CO}} = 1.3450:0.0005:1.3550 \; \text{\AA}$. Some additional points were included for CCSD. For hypofluorous acid (where $\b{g}_\mathrm{scan} = g \, \b{g}$ and $\b{h}_\mathrm{scan} = h \, \b{h}$), $g = -0.010:0.001:0.010$ and $h = -0.1160:0.0016:0.1160$.
Interpolated values are shown in both figures.

The triples excitations used are as follows. In formaldehyde, the excitation is the product of the second-largest singles and the largest doubles excitations in the lower state at $1.3450\; \text{\AA}$. In hypofluorous acid, the excitation is the product of the largest singles and doubles excitations in the lower states at the geometry given by ${R_{\ce{OH}} = 1.1400 \; \text{\AA}}$, $R_{\ce{OF}} = 1.3184\; \text{\AA}$, and $\theta_{\ce{HOF}} = 91.06^\circ$.

\bibliographystyle{aipnum4-1}
\bibliography{library}

\begin{thebibliography}{58}%
\makeatletter
\providecommand \@ifxundefined [1]{%
 \@ifx{#1\undefined}
}%
\providecommand \@ifnum [1]{%
 \ifnum #1\expandafter \@firstoftwo
 \else \expandafter \@secondoftwo
 \fi
}%
\providecommand \@ifx [1]{%
 \ifx #1\expandafter \@firstoftwo
 \else \expandafter \@secondoftwo
 \fi
}%
\providecommand \natexlab [1]{#1}%
\providecommand \enquote  [1]{``#1''}%
\providecommand \bibnamefont  [1]{#1}%
\providecommand \bibfnamefont [1]{#1}%
\providecommand \citenamefont [1]{#1}%
\providecommand \href@noop [0]{\@secondoftwo}%
\providecommand \href [0]{\begingroup \@sanitize@url \@href}%
\providecommand \@href[1]{\@@startlink{#1}\@@href}%
\providecommand \@@href[1]{\endgroup#1\@@endlink}%
\providecommand \@sanitize@url [0]{\catcode `\\12\catcode `\$12\catcode
  `\&12\catcode `\#12\catcode `\^12\catcode `\_12\catcode `\%12\relax}%
\providecommand \@@startlink[1]{}%
\providecommand \@@endlink[0]{}%
\providecommand \url  [0]{\begingroup\@sanitize@url \@url }%
\providecommand \@url [1]{\endgroup\@href {#1}{\urlprefix }}%
\providecommand \urlprefix  [0]{URL }%
\providecommand \Eprint [0]{\href }%
\providecommand \doibase [0]{http://dx.doi.org/}%
\providecommand \selectlanguage [0]{\@gobble}%
\providecommand \bibinfo  [0]{\@secondoftwo}%
\providecommand \bibfield  [0]{\@secondoftwo}%
\providecommand \translation [1]{[#1]}%
\providecommand \BibitemOpen [0]{}%
\providecommand \bibitemStop [0]{}%
\providecommand \bibitemNoStop [0]{.\EOS\space}%
\providecommand \EOS [0]{\spacefactor3000\relax}%
\providecommand \BibitemShut  [1]{\csname bibitem#1\endcsname}%
\let\auto@bib@innerbib\@empty
\bibitem [{\citenamefont {Matsika}\ and\ \citenamefont
  {Krause}(2011)}]{Matsika2011}%
  \BibitemOpen
  \bibfield  {author} {\bibinfo {author} {\bibfnamefont {S.}~\bibnamefont
  {Matsika}}\ and\ \bibinfo {author} {\bibfnamefont {P.}~\bibnamefont
  {Krause}},\ }\href@noop {} {\bibfield  {journal} {\bibinfo  {journal} {Annu.
  Rev. Phys. Chem.}\ }\textbf {\bibinfo {volume} {62}},\ \bibinfo {pages} {621}
  (\bibinfo {year} {2011})}\BibitemShut {NoStop}%
\bibitem [{\citenamefont {Zhu}\ and\ \citenamefont {Yarkony}(2016)}]{Zhu2016}%
  \BibitemOpen
  \bibfield  {author} {\bibinfo {author} {\bibfnamefont {X.}~\bibnamefont
  {Zhu}}\ and\ \bibinfo {author} {\bibfnamefont {D.~R.}\ \bibnamefont
  {Yarkony}},\ }\href@noop {} {\bibfield  {journal} {\bibinfo  {journal} {Mol.
  Phys.}\ }\textbf {\bibinfo {volume} {114}},\ \bibinfo {pages} {1983}
  (\bibinfo {year} {2016})}\BibitemShut {NoStop}%
\bibitem [{\citenamefont {Kim}\ \emph {et~al.}(2015)\citenamefont {Kim},
  \citenamefont {Tao}, \citenamefont {Martinez},\ and\ \citenamefont
  {Bucksbaum}}]{Kim2015}%
  \BibitemOpen
  \bibfield  {author} {\bibinfo {author} {\bibfnamefont {J.}~\bibnamefont
  {Kim}}, \bibinfo {author} {\bibfnamefont {H.}~\bibnamefont {Tao}}, \bibinfo
  {author} {\bibfnamefont {T.~J.}\ \bibnamefont {Martinez}}, \ and\ \bibinfo
  {author} {\bibfnamefont {P.}~\bibnamefont {Bucksbaum}},\ }\href@noop {}
  {\bibfield  {journal} {\bibinfo  {journal} {J. Phys. B}\ }\textbf {\bibinfo
  {volume} {48}},\ \bibinfo {pages} {164003} (\bibinfo {year}
  {2015})}\BibitemShut {NoStop}%
\bibitem [{\citenamefont {Migani}\ \emph {et~al.}(2008)\citenamefont {Migani},
  \citenamefont {Blancafort}, \citenamefont {Robb},\ and\ \citenamefont
  {DeBellis}}]{Migani2008}%
  \BibitemOpen
  \bibfield  {author} {\bibinfo {author} {\bibfnamefont {A.}~\bibnamefont
  {Migani}}, \bibinfo {author} {\bibfnamefont {L.}~\bibnamefont {Blancafort}},
  \bibinfo {author} {\bibfnamefont {M.~A.}\ \bibnamefont {Robb}}, \ and\
  \bibinfo {author} {\bibfnamefont {A.~D.}\ \bibnamefont {DeBellis}},\
  }\href@noop {} {\bibfield  {journal} {\bibinfo  {journal} {J. Am. Chem.
  Soc.}\ }\textbf {\bibinfo {volume} {130}},\ \bibinfo {pages} {6932} (\bibinfo
  {year} {2008})}\BibitemShut {NoStop}%
\bibitem [{\citenamefont {Polli}\ \emph {et~al.}(2010)\citenamefont {Polli},
  \citenamefont {Alto{\`e}}, \citenamefont {Weingart}, \citenamefont
  {Spillane}, \citenamefont {Manzoni}, \citenamefont {Brida}, \citenamefont
  {Tomasello}, \citenamefont {Orlandi}, \citenamefont {Kukura}, \citenamefont
  {Mathies} \emph {et~al.}}]{Polli2010}%
  \BibitemOpen
  \bibfield  {author} {\bibinfo {author} {\bibfnamefont {D.}~\bibnamefont
  {Polli}}, \bibinfo {author} {\bibfnamefont {P.}~\bibnamefont {Alto{\`e}}},
  \bibinfo {author} {\bibfnamefont {O.}~\bibnamefont {Weingart}}, \bibinfo
  {author} {\bibfnamefont {K.~M.}\ \bibnamefont {Spillane}}, \bibinfo {author}
  {\bibfnamefont {C.}~\bibnamefont {Manzoni}}, \bibinfo {author} {\bibfnamefont
  {D.}~\bibnamefont {Brida}}, \bibinfo {author} {\bibfnamefont
  {G.}~\bibnamefont {Tomasello}}, \bibinfo {author} {\bibfnamefont
  {G.}~\bibnamefont {Orlandi}}, \bibinfo {author} {\bibfnamefont
  {P.}~\bibnamefont {Kukura}}, \bibinfo {author} {\bibfnamefont {R.~A.}\
  \bibnamefont {Mathies}},  \emph {et~al.},\ }\href@noop {} {\bibfield
  {journal} {\bibinfo  {journal} {Nature}\ }\textbf {\bibinfo {volume} {467}},\
  \bibinfo {pages} {440} (\bibinfo {year} {2010})}\BibitemShut {NoStop}%
\bibitem [{\citenamefont {Born}\ and\ \citenamefont
  {Oppenheimer}(1927)}]{Born1927}%
  \BibitemOpen
  \bibfield  {author} {\bibinfo {author} {\bibfnamefont {M.}~\bibnamefont
  {Born}}\ and\ \bibinfo {author} {\bibfnamefont {R.}~\bibnamefont
  {Oppenheimer}},\ }\href@noop {} {\bibfield  {journal} {\bibinfo  {journal}
  {Ann. Phys. (Berlin)}\ }\textbf {\bibinfo {volume} {389}},\ \bibinfo {pages}
  {457} (\bibinfo {year} {1927})}\BibitemShut {NoStop}%
\bibitem [{\citenamefont {Ben-Nun}, \citenamefont {Martinez}\ \emph
  {et~al.}(2002)\citenamefont {Ben-Nun}, \citenamefont {Martinez} \emph
  {et~al.}}]{Ben2002}%
  \BibitemOpen
  \bibfield  {author} {\bibinfo {author} {\bibfnamefont {M.}~\bibnamefont
  {Ben-Nun}}, \bibinfo {author} {\bibfnamefont {T.~J.}\ \bibnamefont
  {Martinez}},  \emph {et~al.},\ }\href@noop {} {\bibfield  {journal} {\bibinfo
   {journal} {Adv. Chem. Phys.}\ }\textbf {\bibinfo {volume} {121}},\ \bibinfo
  {pages} {439} (\bibinfo {year} {2002})}\BibitemShut {NoStop}%
\bibitem [{\citenamefont {Stolow}(2003)}]{Stolow2003}%
  \BibitemOpen
  \bibfield  {author} {\bibinfo {author} {\bibfnamefont {A.}~\bibnamefont
  {Stolow}},\ }\href@noop {} {\bibfield  {journal} {\bibinfo  {journal} {Annu.
  Rev. Phys. Chem.}\ }\textbf {\bibinfo {volume} {54}},\ \bibinfo {pages} {89}
  (\bibinfo {year} {2003})}\BibitemShut {NoStop}%
\bibitem [{\citenamefont {Gozem}\ \emph {et~al.}(2014)\citenamefont {Gozem},
  \citenamefont {Melaccio}, \citenamefont {Valentini}, \citenamefont {Filatov},
  \citenamefont {Huix-Rotllant}, \citenamefont {Ferré}, \citenamefont
  {Frutos}, \citenamefont {Angeli}, \citenamefont {Krylov}, \citenamefont
  {Granovsky}, \citenamefont {Lindh},\ and\ \citenamefont
  {Olivucci}}]{Gozem2014}%
  \BibitemOpen
  \bibfield  {author} {\bibinfo {author} {\bibfnamefont {S.}~\bibnamefont
  {Gozem}}, \bibinfo {author} {\bibfnamefont {F.}~\bibnamefont {Melaccio}},
  \bibinfo {author} {\bibfnamefont {A.}~\bibnamefont {Valentini}}, \bibinfo
  {author} {\bibfnamefont {M.}~\bibnamefont {Filatov}}, \bibinfo {author}
  {\bibfnamefont {M.}~\bibnamefont {Huix-Rotllant}}, \bibinfo {author}
  {\bibfnamefont {N.}~\bibnamefont {Ferré}}, \bibinfo {author} {\bibfnamefont
  {L.~M.}\ \bibnamefont {Frutos}}, \bibinfo {author} {\bibfnamefont
  {C.}~\bibnamefont {Angeli}}, \bibinfo {author} {\bibfnamefont {A.~I.}\
  \bibnamefont {Krylov}}, \bibinfo {author} {\bibfnamefont {A.~A.}\
  \bibnamefont {Granovsky}}, \bibinfo {author} {\bibfnamefont {R.}~\bibnamefont
  {Lindh}}, \ and\ \bibinfo {author} {\bibfnamefont {M.}~\bibnamefont
  {Olivucci}},\ }\href@noop {} {\bibfield  {journal} {\bibinfo  {journal} {J.
  Chem. Theory Comput.}\ }\textbf {\bibinfo {volume} {10}},\ \bibinfo {pages}
  {3074} (\bibinfo {year} {2014})}\BibitemShut {NoStop}%
\bibitem [{\citenamefont {Tuna}\ \emph {et~al.}(2015)\citenamefont {Tuna},
  \citenamefont {Lefrancois}, \citenamefont {Łukasz Wolański}, \citenamefont
  {Gozem}, \citenamefont {Schapiro}, \citenamefont {Andruniów}, \citenamefont
  {Dreuw},\ and\ \citenamefont {Olivucci}}]{Tuna2015}%
  \BibitemOpen
  \bibfield  {author} {\bibinfo {author} {\bibfnamefont {D.}~\bibnamefont
  {Tuna}}, \bibinfo {author} {\bibfnamefont {D.}~\bibnamefont {Lefrancois}},
  \bibinfo {author} {\bibnamefont {Łukasz Wolański}}, \bibinfo {author}
  {\bibfnamefont {S.}~\bibnamefont {Gozem}}, \bibinfo {author} {\bibfnamefont
  {I.}~\bibnamefont {Schapiro}}, \bibinfo {author} {\bibfnamefont
  {T.}~\bibnamefont {Andruniów}}, \bibinfo {author} {\bibfnamefont
  {A.}~\bibnamefont {Dreuw}}, \ and\ \bibinfo {author} {\bibfnamefont
  {M.}~\bibnamefont {Olivucci}},\ }\href@noop {} {\bibfield  {journal}
  {\bibinfo  {journal} {J. Chem. Theory Comput.}\ }\textbf {\bibinfo {volume}
  {11}},\ \bibinfo {pages} {5758} (\bibinfo {year} {2015})}\BibitemShut
  {NoStop}%
\bibitem [{\citenamefont {Ben-Nun}, \citenamefont {Quenneville},\ and\
  \citenamefont {Martínez}(2000)}]{Ben2000}%
  \BibitemOpen
  \bibfield  {author} {\bibinfo {author} {\bibfnamefont {M.}~\bibnamefont
  {Ben-Nun}}, \bibinfo {author} {\bibfnamefont {J.}~\bibnamefont
  {Quenneville}}, \ and\ \bibinfo {author} {\bibfnamefont {T.~J.}\ \bibnamefont
  {Martínez}},\ }\href@noop {} {\bibfield  {journal} {\bibinfo  {journal} {J.
  Phys. Chem. A}\ }\textbf {\bibinfo {volume} {104}},\ \bibinfo {pages} {5161}
  (\bibinfo {year} {2000})}\BibitemShut {NoStop}%
\bibitem [{\citenamefont {Bak}\ \emph {et~al.}(1992)\citenamefont {Bak},
  \citenamefont {J{\o}rgensen}, \citenamefont {Jensen}, \citenamefont {Olsen},\
  and\ \citenamefont {Helgaker}}]{Bak1992}%
  \BibitemOpen
  \bibfield  {author} {\bibinfo {author} {\bibfnamefont {K.~L.}\ \bibnamefont
  {Bak}}, \bibinfo {author} {\bibfnamefont {P.}~\bibnamefont {J{\o}rgensen}},
  \bibinfo {author} {\bibfnamefont {H.~J.~A.}\ \bibnamefont {Jensen}}, \bibinfo
  {author} {\bibfnamefont {J.}~\bibnamefont {Olsen}}, \ and\ \bibinfo {author}
  {\bibfnamefont {T.}~\bibnamefont {Helgaker}},\ }\href@noop {} {\bibfield
  {journal} {\bibinfo  {journal} {J. Chem. Phys.}\ }\textbf {\bibinfo {volume}
  {97}},\ \bibinfo {pages} {7573} (\bibinfo {year} {1992})}\BibitemShut
  {NoStop}%
\bibitem [{\citenamefont {Lischka}\ \emph {et~al.}(2004)\citenamefont
  {Lischka}, \citenamefont {Dallos}, \citenamefont {Szalay}, \citenamefont
  {Yarkony},\ and\ \citenamefont {Shepard}}]{Lischka2004}%
  \BibitemOpen
  \bibfield  {author} {\bibinfo {author} {\bibfnamefont {H.}~\bibnamefont
  {Lischka}}, \bibinfo {author} {\bibfnamefont {M.}~\bibnamefont {Dallos}},
  \bibinfo {author} {\bibfnamefont {P.~G.}\ \bibnamefont {Szalay}}, \bibinfo
  {author} {\bibfnamefont {D.~R.}\ \bibnamefont {Yarkony}}, \ and\ \bibinfo
  {author} {\bibfnamefont {R.}~\bibnamefont {Shepard}},\ }\href@noop {}
  {\bibfield  {journal} {\bibinfo  {journal} {J. Chem. Phys.}\ }\textbf
  {\bibinfo {volume} {120}},\ \bibinfo {pages} {7322} (\bibinfo {year}
  {2004})}\BibitemShut {NoStop}%
\bibitem [{\citenamefont {Tajti}\ and\ \citenamefont
  {Szalay}(2009)}]{Tajti2009}%
  \BibitemOpen
  \bibfield  {author} {\bibinfo {author} {\bibfnamefont {A.}~\bibnamefont
  {Tajti}}\ and\ \bibinfo {author} {\bibfnamefont {P.~G.}\ \bibnamefont
  {Szalay}},\ }\href@noop {} {\bibfield  {journal} {\bibinfo  {journal} {J.
  Chem. Phys.}\ }\textbf {\bibinfo {volume} {131}},\ \bibinfo {pages} {124104}
  (\bibinfo {year} {2009})}\BibitemShut {NoStop}%
\bibitem [{\citenamefont {Tew}, \citenamefont {Klopper},\ and\ \citenamefont
  {Helgaker}(2007)}]{Tew2007}%
  \BibitemOpen
  \bibfield  {author} {\bibinfo {author} {\bibfnamefont {D.~P.}\ \bibnamefont
  {Tew}}, \bibinfo {author} {\bibfnamefont {W.}~\bibnamefont {Klopper}}, \ and\
  \bibinfo {author} {\bibfnamefont {T.}~\bibnamefont {Helgaker}},\ }\href@noop
  {} {\bibfield  {journal} {\bibinfo  {journal} {J. Comput. Chem.}\ }\textbf
  {\bibinfo {volume} {28}},\ \bibinfo {pages} {1307} (\bibinfo {year}
  {2007})}\BibitemShut {NoStop}%
\bibitem [{\citenamefont {Coester}(1958)}]{Coester1958}%
  \BibitemOpen
  \bibfield  {author} {\bibinfo {author} {\bibfnamefont {F.}~\bibnamefont
  {Coester}},\ }\href@noop {} {\bibfield  {journal} {\bibinfo  {journal} {Nucl.
  Phys.}\ }\textbf {\bibinfo {volume} {7}},\ \bibinfo {pages} {421} (\bibinfo
  {year} {1958})}\BibitemShut {NoStop}%
\bibitem [{\citenamefont {{\v C}í{\v z}ek}(1966)}]{Cizek1966}%
  \BibitemOpen
  \bibfield  {author} {\bibinfo {author} {\bibfnamefont {J.}~\bibnamefont {{\v
  C}í{\v z}ek}},\ }\href@noop {} {\bibfield  {journal} {\bibinfo  {journal}
  {J. Chem. Phys.}\ }\textbf {\bibinfo {volume} {45}},\ \bibinfo {pages} {4256}
  (\bibinfo {year} {1966})}\BibitemShut {NoStop}%
\bibitem [{\citenamefont {Bartlett}\ and\ \citenamefont
  {Musia\l{}}(2007)}]{Bartlett2007}%
  \BibitemOpen
  \bibfield  {author} {\bibinfo {author} {\bibfnamefont {R.~J.}\ \bibnamefont
  {Bartlett}}\ and\ \bibinfo {author} {\bibfnamefont {M.}~\bibnamefont
  {Musia\l{}}},\ }\href@noop {} {\bibfield  {journal} {\bibinfo  {journal}
  {Rev. Mod. Phys.}\ }\textbf {\bibinfo {volume} {79}},\ \bibinfo {pages} {291}
  (\bibinfo {year} {2007})}\BibitemShut {NoStop}%
\bibitem [{\citenamefont {Lyakh}\ \emph {et~al.}(2012)\citenamefont {Lyakh},
  \citenamefont {Musiał}, \citenamefont {Lotrich},\ and\ \citenamefont
  {Bartlett}}]{Lyakh2012}%
  \BibitemOpen
  \bibfield  {author} {\bibinfo {author} {\bibfnamefont {D.~I.}\ \bibnamefont
  {Lyakh}}, \bibinfo {author} {\bibfnamefont {M.}~\bibnamefont {Musiał}},
  \bibinfo {author} {\bibfnamefont {V.~F.}\ \bibnamefont {Lotrich}}, \ and\
  \bibinfo {author} {\bibfnamefont {R.~J.}\ \bibnamefont {Bartlett}},\
  }\href@noop {} {\bibfield  {journal} {\bibinfo  {journal} {Chem. Rev.}\
  }\textbf {\bibinfo {volume} {112}},\ \bibinfo {pages} {182} (\bibinfo {year}
  {2012})}\BibitemShut {NoStop}%
\bibitem [{\citenamefont {Krylov}(2008)}]{Krylov2008}%
  \BibitemOpen
  \bibfield  {author} {\bibinfo {author} {\bibfnamefont {A.~I.}\ \bibnamefont
  {Krylov}},\ }\href@noop {} {\bibfield  {journal} {\bibinfo  {journal} {Annu.
  Rev. Phys. Chem.}\ }\textbf {\bibinfo {volume} {59}},\ \bibinfo {pages} {433}
  (\bibinfo {year} {2008})}\BibitemShut {NoStop}%
\bibitem [{\citenamefont {Hättig}(2005)}]{Hattig2005}%
  \BibitemOpen
  \bibfield  {author} {\bibinfo {author} {\bibfnamefont {C.}~\bibnamefont
  {Hättig}},\ }in\ \href@noop {} {\emph {\bibinfo {booktitle} {Response Theory
  and Molecular Properties (A Tribute to Jan Linderberg and Poul
  Jørgensen)}}},\ \bibinfo {series} {Adv. Quantum Chem.}, Vol.~\bibinfo
  {volume} {50},\ \bibinfo {editor} {edited by\ \bibinfo {editor}
  {\bibfnamefont {H.}~\bibnamefont {Jensen}}}\ (\bibinfo  {publisher} {Academic
  Press},\ \bibinfo {year} {2005})\ pp.\ \bibinfo {pages} {37 --
  60}\BibitemShut {NoStop}%
\bibitem [{\citenamefont {Purvis~III}\ and\ \citenamefont
  {Bartlett}(1982)}]{Purvis1982}%
  \BibitemOpen
  \bibfield  {author} {\bibinfo {author} {\bibfnamefont {G.~D.}\ \bibnamefont
  {Purvis~III}}\ and\ \bibinfo {author} {\bibfnamefont {R.~J.}\ \bibnamefont
  {Bartlett}},\ }\href@noop {} {\bibfield  {journal} {\bibinfo  {journal} {J.
  Chem. Phys.}\ }\textbf {\bibinfo {volume} {76}},\ \bibinfo {pages} {1910}
  (\bibinfo {year} {1982})}\BibitemShut {NoStop}%
\bibitem [{\citenamefont {Noga}\ and\ \citenamefont
  {Bartlett}(1987)}]{Noga1987}%
  \BibitemOpen
  \bibfield  {author} {\bibinfo {author} {\bibfnamefont {J.}~\bibnamefont
  {Noga}}\ and\ \bibinfo {author} {\bibfnamefont {R.~J.}\ \bibnamefont
  {Bartlett}},\ }\href@noop {} {\bibfield  {journal} {\bibinfo  {journal} {J.
  Chem. Phys.}\ }\textbf {\bibinfo {volume} {86}},\ \bibinfo {pages} {7041}
  (\bibinfo {year} {1987})}\BibitemShut {NoStop}%
\bibitem [{\citenamefont {Noga}\ and\ \citenamefont
  {Bartlett}(1988)}]{Noga1988}%
  \BibitemOpen
  \bibfield  {author} {\bibinfo {author} {\bibfnamefont {J.}~\bibnamefont
  {Noga}}\ and\ \bibinfo {author} {\bibfnamefont {R.~J.}\ \bibnamefont
  {Bartlett}},\ }\href@noop {} {\bibfield  {journal} {\bibinfo  {journal} {J.
  Chem. Phys.}\ }\textbf {\bibinfo {volume} {89}},\ \bibinfo {pages} {3401}
  (\bibinfo {year} {1988})}\BibitemShut {NoStop}%
\bibitem [{\citenamefont {Köhn}\ and\ \citenamefont {Tajti}(2007)}]{Kohn2007}%
  \BibitemOpen
  \bibfield  {author} {\bibinfo {author} {\bibfnamefont {A.}~\bibnamefont
  {Köhn}}\ and\ \bibinfo {author} {\bibfnamefont {A.}~\bibnamefont {Tajti}},\
  }\href@noop {} {\bibfield  {journal} {\bibinfo  {journal} {J. Chem. Phys.}\
  }\textbf {\bibinfo {volume} {127}},\ \bibinfo {pages} {044105} (\bibinfo
  {year} {2007})}\BibitemShut {NoStop}%
\bibitem [{\citenamefont {Christiansen}, \citenamefont {Koch},\ and\
  \citenamefont {Jørgensen}(1995)}]{Christiansen1995}%
  \BibitemOpen
  \bibfield  {author} {\bibinfo {author} {\bibfnamefont {O.}~\bibnamefont
  {Christiansen}}, \bibinfo {author} {\bibfnamefont {H.}~\bibnamefont {Koch}},
  \ and\ \bibinfo {author} {\bibfnamefont {P.}~\bibnamefont {Jørgensen}},\
  }\href@noop {} {\bibfield  {journal} {\bibinfo  {journal} {Chem. Phys.
  Lett.}\ }\textbf {\bibinfo {volume} {243}},\ \bibinfo {pages} {409 }
  (\bibinfo {year} {1995})}\BibitemShut {NoStop}%
\bibitem [{\citenamefont {Plasser}\ \emph {et~al.}(2014)\citenamefont
  {Plasser}, \citenamefont {Crespo-Otero}, \citenamefont {Pederzoli},
  \citenamefont {Pittner}, \citenamefont {Lischka},\ and\ \citenamefont
  {Barbatti}}]{Plasser2014}%
  \BibitemOpen
  \bibfield  {author} {\bibinfo {author} {\bibfnamefont {F.}~\bibnamefont
  {Plasser}}, \bibinfo {author} {\bibfnamefont {R.}~\bibnamefont
  {Crespo-Otero}}, \bibinfo {author} {\bibfnamefont {M.}~\bibnamefont
  {Pederzoli}}, \bibinfo {author} {\bibfnamefont {J.}~\bibnamefont {Pittner}},
  \bibinfo {author} {\bibfnamefont {H.}~\bibnamefont {Lischka}}, \ and\
  \bibinfo {author} {\bibfnamefont {M.}~\bibnamefont {Barbatti}},\ }\href@noop
  {} {\bibfield  {journal} {\bibinfo  {journal} {J. Chem. Theory Comput.}\
  }\textbf {\bibinfo {volume} {10}},\ \bibinfo {pages} {1395} (\bibinfo {year}
  {2014})}\BibitemShut {NoStop}%
\bibitem [{\citenamefont {Andersson}\ \emph {et~al.}(1990)\citenamefont
  {Andersson}, \citenamefont {Malmqvist}, \citenamefont {Roos}, \citenamefont
  {Sadlej},\ and\ \citenamefont {Wolinski}}]{Andersson1990}%
  \BibitemOpen
  \bibfield  {author} {\bibinfo {author} {\bibfnamefont {K.}~\bibnamefont
  {Andersson}}, \bibinfo {author} {\bibfnamefont {P.~A.}\ \bibnamefont
  {Malmqvist}}, \bibinfo {author} {\bibfnamefont {B.~O.}\ \bibnamefont {Roos}},
  \bibinfo {author} {\bibfnamefont {A.~J.}\ \bibnamefont {Sadlej}}, \ and\
  \bibinfo {author} {\bibfnamefont {K.}~\bibnamefont {Wolinski}},\ }\href@noop
  {} {\bibfield  {journal} {\bibinfo  {journal} {J. Phys. Chem.}\ }\textbf
  {\bibinfo {volume} {94}},\ \bibinfo {pages} {5483} (\bibinfo {year}
  {1990})}\BibitemShut {NoStop}%
\bibitem [{\citenamefont {Levine}, \citenamefont {Coe},\ and\ \citenamefont
  {Martínez}(2008)}]{Levine2008}%
  \BibitemOpen
  \bibfield  {author} {\bibinfo {author} {\bibfnamefont {B.~G.}\ \bibnamefont
  {Levine}}, \bibinfo {author} {\bibfnamefont {J.~D.}\ \bibnamefont {Coe}}, \
  and\ \bibinfo {author} {\bibfnamefont {T.~J.}\ \bibnamefont {Martínez}},\
  }\href@noop {} {\bibfield  {journal} {\bibinfo  {journal} {J. Phys. Chem. B}\
  }\textbf {\bibinfo {volume} {112}},\ \bibinfo {pages} {405} (\bibinfo {year}
  {2008})}\BibitemShut {NoStop}%
\bibitem [{\citenamefont {Schirmer}(1982)}]{Schirmer1982}%
  \BibitemOpen
  \bibfield  {author} {\bibinfo {author} {\bibfnamefont {J.}~\bibnamefont
  {Schirmer}},\ }\href@noop {} {\bibfield  {journal} {\bibinfo  {journal}
  {Phys. Rev. A}\ }\textbf {\bibinfo {volume} {26}},\ \bibinfo {pages} {2395}
  (\bibinfo {year} {1982})}\BibitemShut {NoStop}%
\bibitem [{\citenamefont {Dreuw}\ and\ \citenamefont
  {Wormit}(2015)}]{Dreuw2015}%
  \BibitemOpen
  \bibfield  {author} {\bibinfo {author} {\bibfnamefont {A.}~\bibnamefont
  {Dreuw}}\ and\ \bibinfo {author} {\bibfnamefont {M.}~\bibnamefont {Wormit}},\
  }\href@noop {} {\bibfield  {journal} {\bibinfo  {journal} {Wiley Interdiscip.
  Rev. Comput. Mol. Sci.}\ }\textbf {\bibinfo {volume} {5}},\ \bibinfo {pages}
  {82} (\bibinfo {year} {2015})}\BibitemShut {NoStop}%
\bibitem [{\citenamefont {Christiansen}\ \emph {et~al.}(1996)\citenamefont
  {Christiansen}, \citenamefont {Olsen}, \citenamefont {J{\o}rgensen},
  \citenamefont {Koch},\ and\ \citenamefont {Malmqvist}}]{Christiansen1996}%
  \BibitemOpen
  \bibfield  {author} {\bibinfo {author} {\bibfnamefont {O.}~\bibnamefont
  {Christiansen}}, \bibinfo {author} {\bibfnamefont {J.}~\bibnamefont {Olsen}},
  \bibinfo {author} {\bibfnamefont {P.}~\bibnamefont {J{\o}rgensen}}, \bibinfo
  {author} {\bibfnamefont {H.}~\bibnamefont {Koch}}, \ and\ \bibinfo {author}
  {\bibfnamefont {P.-{\AA}.}\ \bibnamefont {Malmqvist}},\ }\href@noop {}
  {\bibfield  {journal} {\bibinfo  {journal} {Chem. Phys. Lett.}\ }\textbf
  {\bibinfo {volume} {261}},\ \bibinfo {pages} {369} (\bibinfo {year}
  {1996})}\BibitemShut {NoStop}%
\bibitem [{\citenamefont {Helgaker}, \citenamefont {J{\o}rgensen},\ and\
  \citenamefont {Olsen}(2014)}]{Helgaker2014}%
  \BibitemOpen
  \bibfield  {author} {\bibinfo {author} {\bibfnamefont {T.}~\bibnamefont
  {Helgaker}}, \bibinfo {author} {\bibfnamefont {P.}~\bibnamefont
  {J{\o}rgensen}}, \ and\ \bibinfo {author} {\bibfnamefont {J.}~\bibnamefont
  {Olsen}},\ }\href@noop {} {\emph {\bibinfo {title} {Molecular
  electronic-structure theory}}}\ (\bibinfo  {publisher} {John Wiley \& Sons},\
  \bibinfo {year} {2014})\BibitemShut {NoStop}%
\bibitem [{\citenamefont {Levine}\ \emph {et~al.}(2006)\citenamefont {Levine},
  \citenamefont {Ko}, \citenamefont {Quenneville},\ and\ \citenamefont
  {Martínez}}]{Levine2006}%
  \BibitemOpen
  \bibfield  {author} {\bibinfo {author} {\bibfnamefont {B.~G.}\ \bibnamefont
  {Levine}}, \bibinfo {author} {\bibfnamefont {C.}~\bibnamefont {Ko}}, \bibinfo
  {author} {\bibfnamefont {J.}~\bibnamefont {Quenneville}}, \ and\ \bibinfo
  {author} {\bibfnamefont {T.~J.}\ \bibnamefont {Martínez}},\ }\href@noop {}
  {\bibfield  {journal} {\bibinfo  {journal} {Mol. Phys.}\ }\textbf {\bibinfo
  {volume} {104}},\ \bibinfo {pages} {1039} (\bibinfo {year}
  {2006})}\BibitemShut {NoStop}%
\bibitem [{\citenamefont {Picconi}\ \emph {et~al.}(2011)\citenamefont
  {Picconi}, \citenamefont {Barone}, \citenamefont {Lami}, \citenamefont
  {Santoro},\ and\ \citenamefont {Improta}}]{Picconi2011}%
  \BibitemOpen
  \bibfield  {author} {\bibinfo {author} {\bibfnamefont {D.}~\bibnamefont
  {Picconi}}, \bibinfo {author} {\bibfnamefont {V.}~\bibnamefont {Barone}},
  \bibinfo {author} {\bibfnamefont {A.}~\bibnamefont {Lami}}, \bibinfo {author}
  {\bibfnamefont {F.}~\bibnamefont {Santoro}}, \ and\ \bibinfo {author}
  {\bibfnamefont {R.}~\bibnamefont {Improta}},\ }\href@noop {} {\bibfield
  {journal} {\bibinfo  {journal} {ChemPhysChem}\ }\textbf {\bibinfo {volume}
  {12}},\ \bibinfo {pages} {1957} (\bibinfo {year} {2011})}\BibitemShut
  {NoStop}%
\bibitem [{\citenamefont {Improta}, \citenamefont {Santoro},\ and\
  \citenamefont {Blancafort}(2016)}]{Improta2016}%
  \BibitemOpen
  \bibfield  {author} {\bibinfo {author} {\bibfnamefont {R.}~\bibnamefont
  {Improta}}, \bibinfo {author} {\bibfnamefont {F.}~\bibnamefont {Santoro}}, \
  and\ \bibinfo {author} {\bibfnamefont {L.}~\bibnamefont {Blancafort}},\
  }\href@noop {} {\bibfield  {journal} {\bibinfo  {journal} {Chemical Reviews}\
  }\textbf {\bibinfo {volume} {116}},\ \bibinfo {pages} {3540} (\bibinfo {year}
  {2016})}\BibitemShut {NoStop}%
\bibitem [{\citenamefont {Hudock}\ \emph {et~al.}(2007)\citenamefont {Hudock},
  \citenamefont {Levine}, \citenamefont {Thompson}, \citenamefont {Satzger},
  \citenamefont {Townsend}, \citenamefont {Gador}, \citenamefont {Ullrich},
  \citenamefont {Stolow},\ and\ \citenamefont {Martínez}}]{Hudock2007}%
  \BibitemOpen
  \bibfield  {author} {\bibinfo {author} {\bibfnamefont {H.~R.}\ \bibnamefont
  {Hudock}}, \bibinfo {author} {\bibfnamefont {B.~G.}\ \bibnamefont {Levine}},
  \bibinfo {author} {\bibfnamefont {A.~L.}\ \bibnamefont {Thompson}}, \bibinfo
  {author} {\bibfnamefont {H.}~\bibnamefont {Satzger}}, \bibinfo {author}
  {\bibfnamefont {D.}~\bibnamefont {Townsend}}, \bibinfo {author}
  {\bibfnamefont {N.}~\bibnamefont {Gador}}, \bibinfo {author} {\bibfnamefont
  {S.}~\bibnamefont {Ullrich}}, \bibinfo {author} {\bibfnamefont
  {A.}~\bibnamefont {Stolow}}, \ and\ \bibinfo {author} {\bibfnamefont {T.~J.}\
  \bibnamefont {Martínez}},\ }\href@noop {} {\bibfield  {journal} {\bibinfo
  {journal} {J. Phys. Chem. A}\ }\textbf {\bibinfo {volume} {111}},\ \bibinfo
  {pages} {8500} (\bibinfo {year} {2007})}\BibitemShut {NoStop}%
\bibitem [{\citenamefont {Asturiol}\ \emph {et~al.}(2009)\citenamefont
  {Asturiol}, \citenamefont {Lasorne}, \citenamefont {Robb},\ and\
  \citenamefont {Blancafort}}]{Asturiol2009}%
  \BibitemOpen
  \bibfield  {author} {\bibinfo {author} {\bibfnamefont {D.}~\bibnamefont
  {Asturiol}}, \bibinfo {author} {\bibfnamefont {B.}~\bibnamefont {Lasorne}},
  \bibinfo {author} {\bibfnamefont {M.~A.}\ \bibnamefont {Robb}}, \ and\
  \bibinfo {author} {\bibfnamefont {L.}~\bibnamefont {Blancafort}},\
  }\href@noop {} {\bibfield  {journal} {\bibinfo  {journal} {J. Phys. Chem. A}\
  }\textbf {\bibinfo {volume} {113}},\ \bibinfo {pages} {10211} (\bibinfo
  {year} {2009})}\BibitemShut {NoStop}%
\bibitem [{\citenamefont {Nakayama}\ \emph {et~al.}(2013)\citenamefont
  {Nakayama}, \citenamefont {Arai}, \citenamefont {Yamazaki},\ and\
  \citenamefont {Taketsugu}}]{Nakayama2013}%
  \BibitemOpen
  \bibfield  {author} {\bibinfo {author} {\bibfnamefont {A.}~\bibnamefont
  {Nakayama}}, \bibinfo {author} {\bibfnamefont {G.}~\bibnamefont {Arai}},
  \bibinfo {author} {\bibfnamefont {S.}~\bibnamefont {Yamazaki}}, \ and\
  \bibinfo {author} {\bibfnamefont {T.}~\bibnamefont {Taketsugu}},\ }\href@noop
  {} {\bibfield  {journal} {\bibinfo  {journal} {J. Chem. Phys.}\ }\textbf
  {\bibinfo {volume} {139}},\ \bibinfo {pages} {214304} (\bibinfo {year}
  {2013})}\BibitemShut {NoStop}%
\bibitem [{\citenamefont {Wolf}\ \emph {et~al.}(2017)\citenamefont {Wolf},
  \citenamefont {Myhre}, \citenamefont {Cryan}, \citenamefont {Coriani},
  \citenamefont {Squibb}, \citenamefont {Battistoni}, \citenamefont {Berrah},
  \citenamefont {Bostedt}, \citenamefont {Bucksbaum}, \citenamefont
  {Coslovich}, \citenamefont {Feifel}, \citenamefont {Gaffney}, \citenamefont
  {Grilj}, \citenamefont {Martínez}, \citenamefont {Miyabe}, \citenamefont
  {Moeller}, \citenamefont {Mucke}, \citenamefont {Natan}, \citenamefont
  {Obaid}, \citenamefont {Osipov}, \citenamefont {Plekan}, \citenamefont
  {Wang}, \citenamefont {Koch},\ and\ \citenamefont {Gühr}}]{Wolf2017}%
  \BibitemOpen
  \bibfield  {author} {\bibinfo {author} {\bibfnamefont {T.}~\bibnamefont
  {Wolf}}, \bibinfo {author} {\bibfnamefont {R.}~\bibnamefont {Myhre}},
  \bibinfo {author} {\bibfnamefont {J.}~\bibnamefont {Cryan}}, \bibinfo
  {author} {\bibfnamefont {S.}~\bibnamefont {Coriani}}, \bibinfo {author}
  {\bibfnamefont {R.}~\bibnamefont {Squibb}}, \bibinfo {author} {\bibfnamefont
  {A.}~\bibnamefont {Battistoni}}, \bibinfo {author} {\bibfnamefont
  {N.}~\bibnamefont {Berrah}}, \bibinfo {author} {\bibfnamefont
  {C.}~\bibnamefont {Bostedt}}, \bibinfo {author} {\bibfnamefont
  {P.}~\bibnamefont {Bucksbaum}}, \bibinfo {author} {\bibfnamefont
  {G.}~\bibnamefont {Coslovich}}, \bibinfo {author} {\bibfnamefont
  {R.}~\bibnamefont {Feifel}}, \bibinfo {author} {\bibfnamefont {K.~J.}\
  \bibnamefont {Gaffney}}, \bibinfo {author} {\bibfnamefont {J.}~\bibnamefont
  {Grilj}}, \bibinfo {author} {\bibfnamefont {T.~J.}\ \bibnamefont
  {Martínez}}, \bibinfo {author} {\bibfnamefont {S.}~\bibnamefont {Miyabe}},
  \bibinfo {author} {\bibfnamefont {S.~P.}\ \bibnamefont {Moeller}}, \bibinfo
  {author} {\bibfnamefont {M.}~\bibnamefont {Mucke}}, \bibinfo {author}
  {\bibfnamefont {A.}~\bibnamefont {Natan}}, \bibinfo {author} {\bibfnamefont
  {R.}~\bibnamefont {Obaid}}, \bibinfo {author} {\bibfnamefont
  {T.}~\bibnamefont {Osipov}}, \bibinfo {author} {\bibfnamefont
  {O.}~\bibnamefont {Plekan}}, \bibinfo {author} {\bibfnamefont
  {S.}~\bibnamefont {Wang}}, \bibinfo {author} {\bibfnamefont {H.}~\bibnamefont
  {Koch}}, \ and\ \bibinfo {author} {\bibfnamefont {M.}~\bibnamefont {Gühr}},\
  }\href@noop {} {\bibfield  {journal} {\bibinfo  {journal} {Nat. Comm.}\
  }\textbf {\bibinfo {volume} {8}} (\bibinfo {year} {2017})}\BibitemShut
  {NoStop}%
\bibitem [{\citenamefont {Kjønstad}\ \emph {et~al.}()\citenamefont
  {Kjønstad}, \citenamefont {Myhre}, \citenamefont {Martínez},\ and\
  \citenamefont {Koch}}]{Kjoenstad2017b}%
  \BibitemOpen
  \bibfield  {author} {\bibinfo {author} {\bibfnamefont {E.~F.}\ \bibnamefont
  {Kjønstad}}, \bibinfo {author} {\bibfnamefont {R.~H.}\ \bibnamefont
  {Myhre}}, \bibinfo {author} {\bibfnamefont {T.~J.}\ \bibnamefont
  {Martínez}}, \ and\ \bibinfo {author} {\bibfnamefont {H.}~\bibnamefont
  {Koch}},\ }\href@noop {} {\bibinfo  {journal} {submitted and will appear on
  arXiv.org}\ }\BibitemShut {NoStop}%
\bibitem [{\citenamefont {{\v C}í{\v z}ek}\ and\ \citenamefont
  {Paldus}(1971)}]{Cizek1971}%
  \BibitemOpen
\bibfield  {journal} {  }\bibfield  {author} {\bibinfo {author} {\bibfnamefont
  {J.}~\bibnamefont {{\v C}í{\v z}ek}}\ and\ \bibinfo {author} {\bibfnamefont
  {J.}~\bibnamefont {Paldus}},\ }\href@noop {} {\bibfield  {journal} {\bibinfo
  {journal} {Int. J. Quant. Chem.}\ }\textbf {\bibinfo {volume} {5}},\ \bibinfo
  {pages} {359} (\bibinfo {year} {1971})}\BibitemShut {NoStop}%
\bibitem [{\citenamefont {Koch}\ and\ \citenamefont
  {J{\o}rgensen}(1990)}]{Koch1990}%
  \BibitemOpen
  \bibfield  {author} {\bibinfo {author} {\bibfnamefont {H.}~\bibnamefont
  {Koch}}\ and\ \bibinfo {author} {\bibfnamefont {P.}~\bibnamefont
  {J{\o}rgensen}},\ }\href@noop {} {\bibfield  {journal} {\bibinfo  {journal}
  {J. Chem. Phys.}\ }\textbf {\bibinfo {volume} {93}},\ \bibinfo {pages} {3333}
  (\bibinfo {year} {1990})}\BibitemShut {NoStop}%
\bibitem [{\citenamefont {Stanton}\ and\ \citenamefont
  {Bartlett}(1993)}]{Stanton1993a}%
  \BibitemOpen
  \bibfield  {author} {\bibinfo {author} {\bibfnamefont {J.~F.}\ \bibnamefont
  {Stanton}}\ and\ \bibinfo {author} {\bibfnamefont {R.~J.}\ \bibnamefont
  {Bartlett}},\ }\href@noop {} {\bibfield  {journal} {\bibinfo  {journal} {J.
  Chem. Phys.}\ }\textbf {\bibinfo {volume} {98}},\ \bibinfo {pages} {7029}
  (\bibinfo {year} {1993})}\BibitemShut {NoStop}%
\bibitem [{\citenamefont {Golub}\ and\ \citenamefont
  {Van~Loan}(2012)}]{Golub2012}%
  \BibitemOpen
  \bibfield  {author} {\bibinfo {author} {\bibfnamefont {G.~H.}\ \bibnamefont
  {Golub}}\ and\ \bibinfo {author} {\bibfnamefont {C.~F.}\ \bibnamefont
  {Van~Loan}},\ }\href@noop {} {\emph {\bibinfo {title} {Matrix
  computations}}},\ \bibinfo {edition} {3rd}\ ed.\ (\bibinfo  {publisher} {JHU
  Press},\ \bibinfo {year} {2012})\BibitemShut {NoStop}%
\bibitem [{\citenamefont {Teller}(1937)}]{Teller1937}%
  \BibitemOpen
  \bibfield  {author} {\bibinfo {author} {\bibfnamefont {E.}~\bibnamefont
  {Teller}},\ }\href@noop {} {\bibfield  {journal} {\bibinfo  {journal} {J.
  Phys. Chem.}\ }\textbf {\bibinfo {volume} {41}},\ \bibinfo {pages} {109}
  (\bibinfo {year} {1937})}\BibitemShut {NoStop}%
\bibitem [{\citenamefont {von Neumann}\ and\ \citenamefont
  {Wigner}(1993)}]{vonNeumann1929}%
  \BibitemOpen
  \bibfield  {author} {\bibinfo {author} {\bibfnamefont {J.}~\bibnamefont {von
  Neumann}}\ and\ \bibinfo {author} {\bibfnamefont {E.}~\bibnamefont
  {Wigner}},\ }in\ \href@noop {} {\emph {\bibinfo {booktitle} {The Collected
  Works of Eugene Paul Wigner}}}\ (\bibinfo  {publisher} {Springer},\ \bibinfo
  {year} {1993})\ pp.\ \bibinfo {pages} {294--297}\BibitemShut {NoStop}%
\bibitem [{\citenamefont {Nanbu}\ and\ \citenamefont
  {Iwata}(1992)}]{Nanbu1992}%
  \BibitemOpen
  \bibfield  {author} {\bibinfo {author} {\bibfnamefont {S.}~\bibnamefont
  {Nanbu}}\ and\ \bibinfo {author} {\bibfnamefont {S.}~\bibnamefont {Iwata}},\
  }\href@noop {} {\bibfield  {journal} {\bibinfo  {journal} {J. Phys. Chem.}\
  }\textbf {\bibinfo {volume} {96}},\ \bibinfo {pages} {2103} (\bibinfo {year}
  {1992})}\BibitemShut {NoStop}%
\bibitem [{\citenamefont {Yarkony}(2001)}]{Yarkony2001b}%
  \BibitemOpen
  \bibfield  {author} {\bibinfo {author} {\bibfnamefont {D.~R.}\ \bibnamefont
  {Yarkony}},\ }\href@noop {} {\bibfield  {journal} {\bibinfo  {journal} {J.
  Phys. Chem. A}\ }\textbf {\bibinfo {volume} {105}},\ \bibinfo {pages} {6277}
  (\bibinfo {year} {2001})}\BibitemShut {NoStop}%
\bibitem [{\citenamefont {Koch}\ \emph {et~al.}(1990)\citenamefont {Koch},
  \citenamefont {Jensen}, \citenamefont {J{\o}rgensen},\ and\ \citenamefont
  {Helgaker}}]{Koch1990b}%
  \BibitemOpen
  \bibfield  {author} {\bibinfo {author} {\bibfnamefont {H.}~\bibnamefont
  {Koch}}, \bibinfo {author} {\bibfnamefont {H.~J.~A.}\ \bibnamefont {Jensen}},
  \bibinfo {author} {\bibfnamefont {P.}~\bibnamefont {J{\o}rgensen}}, \ and\
  \bibinfo {author} {\bibfnamefont {T.}~\bibnamefont {Helgaker}},\ }\href@noop
  {} {\bibfield  {journal} {\bibinfo  {journal} {J. Chem. Phys.}\ }\textbf
  {\bibinfo {volume} {93}},\ \bibinfo {pages} {3345} (\bibinfo {year}
  {1990})}\BibitemShut {NoStop}%
\bibitem [{\citenamefont {Worth}\ and\ \citenamefont
  {Cederbaum}(2004)}]{Worth2004}%
  \BibitemOpen
  \bibfield  {author} {\bibinfo {author} {\bibfnamefont {G.~A.}\ \bibnamefont
  {Worth}}\ and\ \bibinfo {author} {\bibfnamefont {L.~S.}\ \bibnamefont
  {Cederbaum}},\ }\href@noop {} {\bibfield  {journal} {\bibinfo  {journal}
  {Annu. Rev. Phys. Chem.}\ }\textbf {\bibinfo {volume} {55}},\ \bibinfo
  {pages} {127} (\bibinfo {year} {2004})}\BibitemShut {NoStop}%
\bibitem [{\citenamefont {Christiansen}(1999)}]{Christiansen1999}%
  \BibitemOpen
  \bibfield  {author} {\bibinfo {author} {\bibfnamefont {O.}~\bibnamefont
  {Christiansen}},\ }\href@noop {} {\bibfield  {journal} {\bibinfo  {journal}
  {J. Chem. Phys.}\ }\textbf {\bibinfo {volume} {110}},\ \bibinfo {pages} {711}
  (\bibinfo {year} {1999})}\BibitemShut {NoStop}%
\bibitem [{\citenamefont {Stanton}(1993)}]{Stanton1993b}%
  \BibitemOpen
  \bibfield  {author} {\bibinfo {author} {\bibfnamefont {J.~F.}\ \bibnamefont
  {Stanton}},\ }\href@noop {} {\bibfield  {journal} {\bibinfo  {journal} {J.
  Chem. Phys.}\ }\textbf {\bibinfo {volume} {99}},\ \bibinfo {pages} {8840}
  (\bibinfo {year} {1993})}\BibitemShut {NoStop}%
\bibitem [{\citenamefont {Pedersen}\ and\ \citenamefont
  {Koch}(1997)}]{Pedersen1997}%
  \BibitemOpen
  \bibfield  {author} {\bibinfo {author} {\bibfnamefont {T.~B.}\ \bibnamefont
  {Pedersen}}\ and\ \bibinfo {author} {\bibfnamefont {H.}~\bibnamefont
  {Koch}},\ }\href@noop {} {\bibfield  {journal} {\bibinfo  {journal} {J. Chem.
  Phys.}\ }\textbf {\bibinfo {volume} {106}},\ \bibinfo {pages} {8059}
  (\bibinfo {year} {1997})}\BibitemShut {NoStop}%
\bibitem [{\citenamefont {Kobayashi}, \citenamefont {Koch},\ and\ \citenamefont
  {Jørgensen}(1994)}]{Kobayashi1994}%
  \BibitemOpen
  \bibfield  {author} {\bibinfo {author} {\bibfnamefont {R.}~\bibnamefont
  {Kobayashi}}, \bibinfo {author} {\bibfnamefont {H.}~\bibnamefont {Koch}}, \
  and\ \bibinfo {author} {\bibfnamefont {P.}~\bibnamefont {Jørgensen}},\
  }\href@noop {} {\bibfield  {journal} {\bibinfo  {journal} {Chem. Phys.
  Lett.}\ }\textbf {\bibinfo {volume} {219}},\ \bibinfo {pages} {30 } (\bibinfo
  {year} {1994})}\BibitemShut {NoStop}%
\bibitem [{\citenamefont {Aidas}\ \emph {et~al.}(2014)\citenamefont {Aidas},
  \citenamefont {Angeli}, \citenamefont {Bak}, \citenamefont {Bakken},
  \citenamefont {Bast}, \citenamefont {Boman}, \citenamefont {Christiansen},
  \citenamefont {Cimiraglia}, \citenamefont {Coriani}, \citenamefont {Dahle},
  \citenamefont {Dalskov}, \citenamefont {Ekstr\"{o}m}, \citenamefont
  {Enevoldsen}, \citenamefont {Eriksen}, \citenamefont {Ettenhuber},
  \citenamefont {Fern\'{a}ndez}, \citenamefont {Ferrighi}, \citenamefont
  {Fliegl}, \citenamefont {Frediani}, \citenamefont {Hald}, \citenamefont
  {Halkier}, \citenamefont {H\"{a}ttig}, \citenamefont {Heiberg}, \citenamefont
  {Helgaker}, \citenamefont {Hennum}, \citenamefont {Hettema}, \citenamefont
  {Hjerten\ae{}s}, \citenamefont {H\o{}st}, \citenamefont {H\o{}yvik},
  \citenamefont {Iozzi}, \citenamefont {Jans\'{i}k}, \citenamefont {Jensen},
  \citenamefont {Jonsson}, \citenamefont {J\o{}rgensen}, \citenamefont
  {Kauczor}, \citenamefont {Kirpekar}, \citenamefont {Kj\ae{}rgaard},
  \citenamefont {Klopper}, \citenamefont {Knecht}, \citenamefont {Kobayashi},
  \citenamefont {Koch}, \citenamefont {Kongsted}, \citenamefont {Krapp},
  \citenamefont {Kristensen}, \citenamefont {Ligabue}, \citenamefont
  {Lutn\ae{}s}, \citenamefont {Melo}, \citenamefont {Mikkelsen}, \citenamefont
  {Myhre}, \citenamefont {Neiss}, \citenamefont {Nielsen}, \citenamefont
  {Norman}, \citenamefont {Olsen}, \citenamefont {Olsen}, \citenamefont
  {Osted}, \citenamefont {Packer}, \citenamefont {Pawlowski}, \citenamefont
  {Pedersen}, \citenamefont {Provasi}, \citenamefont {Reine}, \citenamefont
  {Rinkevicius}, \citenamefont {Ruden}, \citenamefont {Ruud}, \citenamefont
  {Rybkin}, \citenamefont {Sa\l{}ek}, \citenamefont {Samson}, \citenamefont
  {de~Mer\'{a}s}, \citenamefont {Saue}, \citenamefont {Sauer}, \citenamefont
  {Schimmelpfennig}, \citenamefont {Sneskov}, \citenamefont {Steindal},
  \citenamefont {Sylvester-Hvid}, \citenamefont {Taylor}, \citenamefont
  {Teale}, \citenamefont {Tellgren}, \citenamefont {Tew}, \citenamefont
  {Thorvaldsen}, \citenamefont {Th\o{}gersen}, \citenamefont {Vahtras},
  \citenamefont {Watson}, \citenamefont {Wilson}, \citenamefont {Ziolkowski},\
  and\ \citenamefont {\AA{}gren}}]{DALTON}%
  \BibitemOpen
  \bibfield  {author} {\bibinfo {author} {\bibfnamefont {K.}~\bibnamefont
  {Aidas}}, \bibinfo {author} {\bibfnamefont {C.}~\bibnamefont {Angeli}},
  \bibinfo {author} {\bibfnamefont {K.~L.}\ \bibnamefont {Bak}}, \bibinfo
  {author} {\bibfnamefont {V.}~\bibnamefont {Bakken}}, \bibinfo {author}
  {\bibfnamefont {R.}~\bibnamefont {Bast}}, \bibinfo {author} {\bibfnamefont
  {L.}~\bibnamefont {Boman}}, \bibinfo {author} {\bibfnamefont
  {O.}~\bibnamefont {Christiansen}}, \bibinfo {author} {\bibfnamefont
  {R.}~\bibnamefont {Cimiraglia}}, \bibinfo {author} {\bibfnamefont
  {S.}~\bibnamefont {Coriani}}, \bibinfo {author} {\bibfnamefont
  {P.}~\bibnamefont {Dahle}}, \bibinfo {author} {\bibfnamefont {E.~K.}\
  \bibnamefont {Dalskov}}, \bibinfo {author} {\bibfnamefont {U.}~\bibnamefont
  {Ekstr\"{o}m}}, \bibinfo {author} {\bibfnamefont {T.}~\bibnamefont
  {Enevoldsen}}, \bibinfo {author} {\bibfnamefont {J.~J.}\ \bibnamefont
  {Eriksen}}, \bibinfo {author} {\bibfnamefont {P.}~\bibnamefont {Ettenhuber}},
  \bibinfo {author} {\bibfnamefont {B.}~\bibnamefont {Fern\'{a}ndez}}, \bibinfo
  {author} {\bibfnamefont {L.}~\bibnamefont {Ferrighi}}, \bibinfo {author}
  {\bibfnamefont {H.}~\bibnamefont {Fliegl}}, \bibinfo {author} {\bibfnamefont
  {L.}~\bibnamefont {Frediani}}, \bibinfo {author} {\bibfnamefont
  {K.}~\bibnamefont {Hald}}, \bibinfo {author} {\bibfnamefont {A.}~\bibnamefont
  {Halkier}}, \bibinfo {author} {\bibfnamefont {C.}~\bibnamefont {H\"{a}ttig}},
  \bibinfo {author} {\bibfnamefont {H.}~\bibnamefont {Heiberg}}, \bibinfo
  {author} {\bibfnamefont {T.}~\bibnamefont {Helgaker}}, \bibinfo {author}
  {\bibfnamefont {A.~C.}\ \bibnamefont {Hennum}}, \bibinfo {author}
  {\bibfnamefont {H.}~\bibnamefont {Hettema}}, \bibinfo {author} {\bibfnamefont
  {E.}~\bibnamefont {Hjerten\ae{}s}}, \bibinfo {author} {\bibfnamefont
  {S.}~\bibnamefont {H\o{}st}}, \bibinfo {author} {\bibfnamefont {I.-M.}\
  \bibnamefont {H\o{}yvik}}, \bibinfo {author} {\bibfnamefont {M.~F.}\
  \bibnamefont {Iozzi}}, \bibinfo {author} {\bibfnamefont {B.}~\bibnamefont
  {Jans\'{i}k}}, \bibinfo {author} {\bibfnamefont {H.~J.~{\relax Aa}.}\
  \bibnamefont {Jensen}}, \bibinfo {author} {\bibfnamefont {D.}~\bibnamefont
  {Jonsson}}, \bibinfo {author} {\bibfnamefont {P.}~\bibnamefont
  {J\o{}rgensen}}, \bibinfo {author} {\bibfnamefont {J.}~\bibnamefont
  {Kauczor}}, \bibinfo {author} {\bibfnamefont {S.}~\bibnamefont {Kirpekar}},
  \bibinfo {author} {\bibfnamefont {T.}~\bibnamefont {Kj\ae{}rgaard}}, \bibinfo
  {author} {\bibfnamefont {W.}~\bibnamefont {Klopper}}, \bibinfo {author}
  {\bibfnamefont {S.}~\bibnamefont {Knecht}}, \bibinfo {author} {\bibfnamefont
  {R.}~\bibnamefont {Kobayashi}}, \bibinfo {author} {\bibfnamefont
  {H.}~\bibnamefont {Koch}}, \bibinfo {author} {\bibfnamefont {J.}~\bibnamefont
  {Kongsted}}, \bibinfo {author} {\bibfnamefont {A.}~\bibnamefont {Krapp}},
  \bibinfo {author} {\bibfnamefont {K.}~\bibnamefont {Kristensen}}, \bibinfo
  {author} {\bibfnamefont {A.}~\bibnamefont {Ligabue}}, \bibinfo {author}
  {\bibfnamefont {O.~B.}\ \bibnamefont {Lutn\ae{}s}}, \bibinfo {author}
  {\bibfnamefont {J.~I.}\ \bibnamefont {Melo}}, \bibinfo {author}
  {\bibfnamefont {K.~V.}\ \bibnamefont {Mikkelsen}}, \bibinfo {author}
  {\bibfnamefont {R.~H.}\ \bibnamefont {Myhre}}, \bibinfo {author}
  {\bibfnamefont {C.}~\bibnamefont {Neiss}}, \bibinfo {author} {\bibfnamefont
  {C.~B.}\ \bibnamefont {Nielsen}}, \bibinfo {author} {\bibfnamefont
  {P.}~\bibnamefont {Norman}}, \bibinfo {author} {\bibfnamefont
  {J.}~\bibnamefont {Olsen}}, \bibinfo {author} {\bibfnamefont {J.~M.~H.}\
  \bibnamefont {Olsen}}, \bibinfo {author} {\bibfnamefont {A.}~\bibnamefont
  {Osted}}, \bibinfo {author} {\bibfnamefont {M.~J.}\ \bibnamefont {Packer}},
  \bibinfo {author} {\bibfnamefont {F.}~\bibnamefont {Pawlowski}}, \bibinfo
  {author} {\bibfnamefont {T.~B.}\ \bibnamefont {Pedersen}}, \bibinfo {author}
  {\bibfnamefont {P.~F.}\ \bibnamefont {Provasi}}, \bibinfo {author}
  {\bibfnamefont {S.}~\bibnamefont {Reine}}, \bibinfo {author} {\bibfnamefont
  {Z.}~\bibnamefont {Rinkevicius}}, \bibinfo {author} {\bibfnamefont {T.~A.}\
  \bibnamefont {Ruden}}, \bibinfo {author} {\bibfnamefont {K.}~\bibnamefont
  {Ruud}}, \bibinfo {author} {\bibfnamefont {V.~V.}\ \bibnamefont {Rybkin}},
  \bibinfo {author} {\bibfnamefont {P.}~\bibnamefont {Sa\l{}ek}}, \bibinfo
  {author} {\bibfnamefont {C.~C.~M.}\ \bibnamefont {Samson}}, \bibinfo {author}
  {\bibfnamefont {A.~S.}\ \bibnamefont {de~Mer\'{a}s}}, \bibinfo {author}
  {\bibfnamefont {T.}~\bibnamefont {Saue}}, \bibinfo {author} {\bibfnamefont
  {S.~P.~A.}\ \bibnamefont {Sauer}}, \bibinfo {author} {\bibfnamefont
  {B.}~\bibnamefont {Schimmelpfennig}}, \bibinfo {author} {\bibfnamefont
  {K.}~\bibnamefont {Sneskov}}, \bibinfo {author} {\bibfnamefont {A.~H.}\
  \bibnamefont {Steindal}}, \bibinfo {author} {\bibfnamefont {K.~O.}\
  \bibnamefont {Sylvester-Hvid}}, \bibinfo {author} {\bibfnamefont {P.~R.}\
  \bibnamefont {Taylor}}, \bibinfo {author} {\bibfnamefont {A.~M.}\
  \bibnamefont {Teale}}, \bibinfo {author} {\bibfnamefont {E.~I.}\ \bibnamefont
  {Tellgren}}, \bibinfo {author} {\bibfnamefont {D.~P.}\ \bibnamefont {Tew}},
  \bibinfo {author} {\bibfnamefont {A.~J.}\ \bibnamefont {Thorvaldsen}},
  \bibinfo {author} {\bibfnamefont {L.}~\bibnamefont {Th\o{}gersen}}, \bibinfo
  {author} {\bibfnamefont {O.}~\bibnamefont {Vahtras}}, \bibinfo {author}
  {\bibfnamefont {M.~A.}\ \bibnamefont {Watson}}, \bibinfo {author}
  {\bibfnamefont {D.~J.~D.}\ \bibnamefont {Wilson}}, \bibinfo {author}
  {\bibfnamefont {M.}~\bibnamefont {Ziolkowski}}, \ and\ \bibinfo {author}
  {\bibfnamefont {H.}~\bibnamefont {\AA{}gren}},\ }\href@noop {} {\bibfield
  {journal} {\bibinfo  {journal} {WIREs Comput.~Mol.~Sci.}\ }\textbf {\bibinfo
  {volume} {4}},\ \bibinfo {pages} {269} (\bibinfo {year} {2014})}\BibitemShut
  {NoStop}%
\bibitem [{\citenamefont {Koch}\ \emph {et~al.}(1997)\citenamefont {Koch},
  \citenamefont {Christiansen}, \citenamefont {Jo}, \citenamefont
  {de~Mer{\'a}s}, \citenamefont {Helgaker} \emph {et~al.}}]{Koch1997}%
  \BibitemOpen
  \bibfield  {author} {\bibinfo {author} {\bibfnamefont {H.}~\bibnamefont
  {Koch}}, \bibinfo {author} {\bibfnamefont {O.}~\bibnamefont {Christiansen}},
  \bibinfo {author} {\bibfnamefont {P.}~\bibnamefont {Jo}}, \bibinfo {author}
  {\bibfnamefont {A.~M.~S.}\ \bibnamefont {de~Mer{\'a}s}}, \bibinfo {author}
  {\bibfnamefont {T.}~\bibnamefont {Helgaker}},  \emph {et~al.},\ }\href@noop
  {} {\bibfield  {journal} {\bibinfo  {journal} {J. Chem. Phys.}\ }\textbf
  {\bibinfo {volume} {106}},\ \bibinfo {pages} {1808} (\bibinfo {year}
  {1997})}\BibitemShut {NoStop}%
\bibitem [{\citenamefont {Stanton}\ \emph {et~al.}(2009)\citenamefont
  {Stanton}, \citenamefont {Gauss}, \citenamefont {Harding}, \citenamefont
  {Szalay}, \citenamefont {Auer}, \citenamefont {Bartlett}, \citenamefont
  {Benedikt}, \citenamefont {Berger}, \citenamefont {Bernholdt}, \citenamefont
  {Bomble} \emph {et~al.}}]{cfour}%
  \BibitemOpen
  \bibfield  {author} {\bibinfo {author} {\bibfnamefont {J.}~\bibnamefont
  {Stanton}}, \bibinfo {author} {\bibfnamefont {J.}~\bibnamefont {Gauss}},
  \bibinfo {author} {\bibfnamefont {M.}~\bibnamefont {Harding}}, \bibinfo
  {author} {\bibfnamefont {P.}~\bibnamefont {Szalay}}, \bibinfo {author}
  {\bibfnamefont {A.}~\bibnamefont {Auer}}, \bibinfo {author} {\bibfnamefont
  {R.}~\bibnamefont {Bartlett}}, \bibinfo {author} {\bibfnamefont
  {U.}~\bibnamefont {Benedikt}}, \bibinfo {author} {\bibfnamefont
  {C.}~\bibnamefont {Berger}}, \bibinfo {author} {\bibfnamefont
  {D.}~\bibnamefont {Bernholdt}}, \bibinfo {author} {\bibfnamefont
  {Y.}~\bibnamefont {Bomble}},  \emph {et~al.},\ }\href@noop {} {\bibfield
  {journal} {\bibinfo  {journal} {For the current version, see
  http://www.cfour.de}\ } (\bibinfo {year} {2009})}\BibitemShut {NoStop}%
\end{thebibliography}%

\section*{Acknowledgements} We thank Rolf H.~Myhre and Todd J.~Martínez for enlightening discussions during the preparation of the manuscript. We also wish to thank Thomas Wolf for commenting on a draft of the manuscript. Computer resources from NOTUR project nn2962k are acknowledged. Henrik Koch acknowledges financial support from the FP7-PEOPLE-2013-IOF funding scheme (Project No. 625321). Partial support for this work was provided by the AMOS program within the Chemical Sciences, Geosciences, and Biosciences Division of the Office of Basic Energy Sciences, Office of Science, US Department of Energy. We further acknowledge support from the Norwegian Research Council through FRINATEK project no. 263110/F20.

\section*{Author contributions} The authors contributed equally to all parts of the paper.

\section*{Competing financial interests} The authors declare no competing financial interests.

\section*{Supporting Information}
\subsection*{Derivation of the generalized orthogonality in the full space limit}
The energies of the coupled cluster model equal the eigenvalues of the matrix
\begin{align}
(\J + E_0 \, \b{I})_{\mu\nu} = \Tbraket{\Phi_\mu}{e^{-T} \, H \, e^T}{\Phi_\nu}, \quad E_0 = \Tbraket{\Phi_0}{e^{-T} \, H \, e^T}{\Phi_0},
\end{align}
where $T$ has been determined from the amplitude equations (see, \emph{e.g.}, Purvis and Bartlett\citep{Purvis1982}), and where $\mu,\nu \geq 0$. In the limit where $\mu$ and $\nu$ runs over all excitations, the identity operator can be written $\ident = \sum_{\mu \geq 0} \ket{\Phi_\mu}\bra{\Phi_\mu}$, where we assume that $\{ \ket{\Phi_\mu} \}_\mu$ is an orthonormal basis. Inserting this resolution of $\ident$ before and after $H$ leads to the following expression for $\J + E_0 \, \b{I}$: 
\begin{align}
\begin{split}
(\J + E_0 \, \b{I})_{\mu\nu} &= \sum_{\tau \sigma \geq 0} \Tbraket{\Phi_\mu}{e^{-T}}{\Phi_\tau}\Tbraket{\Phi_\tau}{H}{\Phi_\sigma}\Tbraket{\Phi_\sigma}{e^{T}}{\Phi_\nu} \\
&= \sum_{\tau \sigma \geq 0} Q_{\mu\tau}^{-1} \, H_{\tau\sigma} \, Q_{\sigma \nu} \\
&= (\b{Q}^{-1} \b{H} \, \b{Q})_{\mu\nu}. \label{eq:similarity}
\end{split}
\end{align}
That $\b{Q}^{-1}$ is the inverse of $\b{Q}$ is straightforwardly verified:
\begin{align}
(\b{Q}^{-1} \b{Q})_{\mu\nu} = \sum_{\tau \geq 0} \Tbraket{\Phi_\mu}{e^{-T}}{\Phi_\tau} \Tbraket{\Phi_\tau}{e^{T}}{\Phi_\nu} = \Tbraket{\Phi_\mu}{e^{-T} e^{T}}{\Phi_\nu} = \delta_{\mu\nu}.
\end{align}
It follows from equation \eqref{eq:similarity} that the left and right eigenvectors of $\J + E_0 \, \b{I}$, and therefore of $\J$, satisfy (up to normalization)
\begin{align}
\b{l}_k^T (\b{Q}^T \b{Q})^{-1} \b{l}_l = \delta_{kl}, \quad \b{r}_k^T \b{Q}^T \b{Q} \, \b{r}_l = \delta_{kl},
\end{align}
because the eigenvectors of $\b{H}$ are orthogonal due to its Hermiticity.

\subsection*{Implementation}
The model was implemented in a local version of the Dalton quantum chemistry suite\citep{DALTON}, and a local coupled cluster program currently in the initial phases of development.

\subsubsection*{The projection vector and the coupled cluster Jacobian}

In similarity constrained CCSD (SCCSD), an additional triple excitation and amplitude is added to the cluster operator, $\mathcal{T}$. With this cluster operator, both the projection vector, ${\Omega_\mu = \Tbraket{\Phi_\mu}{e^{-\mathcal{T}} H \, e^\mathcal{T}}{\Phi_0}}$, and the Jacobian matrix, $A_{\mu\nu} = \Tbraket{\Phi_\mu}{e^{-\mathcal{T}} [H,\tau_\nu] \, e^\mathcal{T}}{\Phi_0}$, are modified relative to CCSD. We refer to the literature for detailed expressions of $\bsym{\Omega}$ and $\J$ in CCSD\citep{Purvis1982,Koch1990}.

First we introduce a spin-adapted biorthonormal excitation manifold in which to express our matrices. We use the elementary basis (see p.~691-692 in Helgaker \emph{et al}.\citep{Helgaker2014}) 
\begin{align}
\Ket{\sg{i}{a}} = E_{ai} \, \ket{\Phi_0}, \quad \Ket{\db{i}{a}{j}{b}} = E_{ai} \, E_{bj} \, \ket{\Phi_0}
\end{align}
as the right basis, where $E_{ai} = \cre{a\alpha}\ann{i\alpha} + \cre{a\beta}\ann{i\beta}$, and
\begin{align}
\Bra{\sg{i}{a}} = \frac{1}{2} \bra{\Phi_0}\, E_{ai}\adj, \quad \Bra{\db{i}{a}{j}{b}} = \frac{1}{1 + \delta_{ai,bj}} \Bigl( \frac{1}{3} \bra{\Phi_0} \, E_{ai}\adj E_{bj}\adj + \frac{1}{6} \, E_{aj}\adj E_{bi}\adj  \Bigr)
\end{align}
as the left basis.

We may now list explicit expressions for the projection vector and the transformation by the coupled cluster Jacobian, as well as its transpose: $\bsym{\Omega}$, $\bsym{\rho} = \J \, \b{c}$, and $\bsym{\sigma} = \J^T \, \b{b}$. Let $T = T_1 + T_2 + T_3$, where $T_3$ is general for now:
\begin{align}
T_1 = \sum_{ai} t_i^a \, E_{ai}, \;\; T_2 = \frac{1}{2} \sum_{aibj} t_{ij}^{ab} \, E_{ai} \, E_{bj}, \;\; T_3 = \frac{1}{6} \sum_{aibjck} t_{ijk}^{abc} \, E_{ai} \, E_{bj} \, E_{ck}.
\end{align}
For $\bsym{\rho} = \J \, \b{c}$ we have $\rho_{ai} = \rho_{ai}^\mathrm{CCSD}$ and
\begin{align}
\begin{split}
\rho_{aibj} - \rho_{aibj}^\mathrm{CCSD} = \frac{1}{1 + \delta_{ai,bj}}  \mathscr{P}_{ij}^{ab}  \Bigl( \sum_{ck} (t_{ijk}^{abc} - t_{ikj}^{abc}) & X_{kc} - \sum_{ckl} (2 \, t_{jkl}^{bac} - t_{lkj}^{bac} - t_{jlk}^{bac})Y_{lcki} \\
&\quad+ \sum_{cdk} (2 \, t_{jik}^{bcd} - t_{kij}^{bcd} - t_{jki}^{bcd}) Z_{ackd}\Bigr),
\end{split} 
\end{align}
where
\begin{align}
X_{ld} = \sum_{ck} L_{kcld} \, c_{ck}, \;\; Y_{kclj} = \sum_d g_{kcld} \, c_{dj}, \;\; Z_{acld} &= -\sum_k g_{kcld} \, c_{ak}.
\end{align}
In the above, we have introduced $\mathscr{P}_{ij}^{ab}$, whose effect is to add all permutations of the index pairs $(ai)$ and $(bj)$, and $g_{pqrs}$, the two-electron $T_1$ transformed integrals associated with the $T_1$ transformed Hamiltonian $\hat{H}$. The projection vector has the singles contribution
\begin{align}
\Omega_{ai}-\Omega_{ai}^\mathrm{CCSD} = \sum_{bjck} (t_{ijk}^{abc} - t_{ijk}^{cba}) \, L_{jbkc}, \quad L_{jbkc} = 2 \, g_{jbkc} - g_{jckb}.
\end{align}
The doubles contribution is identical to that of $\bsym{\rho}$, except that the $X$, $Y$, and $Z$ intermediates assume the altered definitions
\begin{align}
X_{kc} = F_{kc}, \;\; Y_{lcki} = g_{lcki}, \;\; Z_{ackd} = g_{ackd}.
\end{align}
We have denoted by $F_{pq}$ the elements of the $T_1$ transformed Fock operator, which is defined as the ordinary Fock operator but with $T_1$ transformed integrals. For $\bsym{\sigma} = \J^T \, \b{b}$ we have $\sigma_{aibj} = \sigma_{aibj}^\mathrm{CCSD}$ and, finally,
\begin{align}
\begin{split}
\sigma_{ck}-\sigma_{ck}^\mathrm{CCSD} = \sum_{dlemfn} & (t_{lmn}^{def} - t_{lnm}^{def}) \, b_{dlem} \,  L_{kcnf} \\
&+ \sum_{dlemfn} (t_{mln}^{def}+t_{lnm}^{def}-2\,t_{lmn}^{def}) \, b_{dlcn} \, g_{mekf} \\
&+ \sum_{dlemfn} (t_{lnm}^{def}+t_{nml}^{def}-2\,t_{lmn}^{def}) \, b_{dlek} \, g_{mcnf}.
\end{split}
\end{align}
The expressions listed above are those of coupled cluster singles doubles triples (CCSDT) and are also found in the literature\citep{Koch1997}. In the simiarity constrained formalism, one particular triple excitation ($\tau_{IJK}^{ABC}$) is selected to be non-zero. We can therefore write
\begin{align}
t_{ijk}^{abc} = \zeta \, \mathscr{P}_{IJK}^{ABC} \delta_{aibjck,AIBJCK},
\end{align}
where $\zeta$ is the magnitude of the chosen triple amplitude, and substitute this $t_{ijk}^{abc}$ in the above expressions. Here we have introduced $\mathscr{P}_{IJK}^{ABC}$, which permutes the index pairs $(AI)$, $(BJ)$, and $(CK)$. Doing the substitution results in
\begin{align}
\begin{split}
\Omega_{ai} - \Omega_{ai}^{\mathrm{\scalebox{0.7}{CCSD}}} = \zeta \, \mathscr{P}_{IJK}^{ABC} (  \delta_{ai,AI} \, L_{JBKC}  - \delta_{ai,CI} \, L_{JBKA})
\end{split} \label{eq:omega_singles}
\end{align}
for the singles part of $\bsym{\Omega}$. The doubles part $\bsym{\Omega}$ is equal to that of $\bsym{\rho}$ (with redefined $X$, $Y$, and $Z$). For $\bsym{\rho}$, we find $\rho_{ai} = 0$ and
\begin{align}
\begin{split}
\rho_{aibj}-\rho_{aibj}^\mathrm{CCSD} = \frac{\zeta}{1 + \delta_{ai,bj}} \mathscr{P}_{ij}^{ab} & \mathscr{P}_{IJK}^{ABC}     \Bigl( \delta_{aibj}^{AIBJ} X_{KC} - \delta_{aibj}^{AIBK} X_{JC} \\
&\;\;-(2 \, \delta_{bja}^{AIB} Y_{KCJi}- \delta_{baj}^{ABK} Y_{ICJi} -\delta_{bja}^{AIB} Y_{JCKi} ) \\
&\;\;+  2 \, \delta_{bji}^{AIJ} Z_{aBKC} - \delta_{bij}^{AJK} Z_{aBIC} - \delta_{bji}^{AIK} Z_{aBJC}  \Bigr).
\end{split} \label{eq:right}
\end{align}
Finally, for $\bsym{\sigma}$ we find $\sigma_{aibj} = \sigma_{aibj}^\mathrm{CCSD}$ and
\begin{align}
\begin{split}
\sigma_{ck}-\sigma_{ck}^\mathrm{CCSD} = \zeta \, P_{IJK}^{ABC} & (b_{AIBJ} L_{kcKC} - b_{AIBK} L_{kcJC} \\
&+ b_{AJcK}g_{IBkC} + b_{AIcJ}g_{KBkC} - 2 \, b_{AIcK}g_{JBkC} \\
&+ b_{AIBk} g_{KcJC} + b_{AKBk} g_{JcIC} - 2 \, b_{AIBk} g_{JcKC}). \label{eq:left}
\end{split}
\end{align}
The equations we have implemented are \eqref{eq:omega_singles}, for $\bsym{\Omega}_1$, \eqref{eq:right}, for $\bsym{\Omega}_2$ and $\bsym{\rho}$, and \eqref{eq:left}, for $\bsym{\sigma}$. For completeness, we note that expressions for energy $E$ and the $\bsym{\eta}$ vector are unchanged:
\begin{align}
E &= \Tbraket{\Phi_0}{e^{-\mathcal{T}}H \, e^\mathcal{T}}{\Phi_0} = E^\mathrm{CCSD}, \\ \eta_\nu &= \Tbraket{\Phi_0}{e^{-\mathcal{T}} [H,\tau_\nu] \, e^\mathcal{T}}{\Phi_0} = \eta_\nu^\mathrm{CCSD}.
\end{align}

\begin{figure*}
\begin{minipage}{\linewidth}
\begin{algorithm}[H]
  \caption{The SCCSD algorithm}
  \label{EPSA}
   \begin{algorithmic}[1]
   \State Select two states $k$ and $l$ and a triple excitation $\tau_{IJK}^{ABC}$.
   \State Set $t_{IJK}^{ABC} = 0$.
   \State For the given $t_{IJK}^{ABC}$, solve $\Omega_{\mu_1} = 0$ and $\Omega_{\mu_2} = 0$ for $t_{\mu_1}$ and $t_{\mu_2}$. 
   \State Solve the multiplier equation $\b{\overline{t}}^T \J = - \veta^T$ for the multipliers $\b{\overline{t}}$.
   \State Solve the eigenvalue equation $\J \, \b{r}_i = \omega_i \, \b{r}_i$ for the excited states $\b{r}_i$.
   \State Evaluate the generalized overlap $f(\mathcal{T})$.
   \If{$f(\mathcal{T}) = 0$} 
   \State Stop.
   \Else
   \State Estimate $\partial f(\mathcal{T})/\partial t_{IJK}^{ABC}$ by numerical differentiation.
   \State Perform a Newton-Raphson step: $t_{IJK}^{ABC} = t_{IJK}^{ABC} - f(\mathcal{T})/(\partial f(\mathcal{T})/\partial t_{IJK}^{ABC})$.
   \State Go to 3.
   \EndIf
   \end{algorithmic}
\end{algorithm}
\end{minipage}
\end{figure*}

\subsubsection*{The generalized overlap} For notational purposes, we denote full space quantities (\emph{i.e.}, with $\mu,\nu \geq 0$) by caligraphic font, $\mathcal{X}$, reserving $X$ for the excited-excited block ($\mu, \nu > 0$). Then we can write
\begin{align}
\bsym{\mathcal{A}} = \begin{pmatrix} 0 & \veta^T \\ 0 & \J \end{pmatrix}
\end{align}
and
\begin{align}
\bsym{\mathcal{Q}} = \begin{pmatrix} 1 & 0 \\ \vq & \b{Q} \end{pmatrix}, \quad q_\mu = \Tbraket{\Phi_\mu}{e^\mathcal{T}}{\Phi_0}, \quad Q_{\mu\nu} = \Tbraket{\Phi_\mu}{e^\mathcal{T}}{\Phi_\nu}.
\end{align}
Note that since we use a biorthonormal basis for $\bsym{\mathcal{Q}}$ and $\bsym{\mathcal{A}}$, the elementary overlap matrix $\bsym{\mathcal{S}}$ (the overlap of the $\ket{\Phi_\mu}$) enters the expression for the generalized overlap $f$:
\begin{align}
\bsym{\mathcal{A}} + E_0 \; \b{I} = \bsym{\mathcal{Q}}^{-1} \bsym{\mathcal{S}}^{-1} \bsym{\mathcal{H}} \; \bsym{\mathcal{Q}},
\end{align}
where $\bsym{\mathcal{H}}$ is $H$ expressed in the elementary basis (the kets $\ket{\Phi_\mu}$). With this notation, the generalized overlap $f$ between the state vectors $\bsym{\mathcal{R}}_k$ and $\bsym{\mathcal{R}}_l$ reads
\begin{align}
f(\mathcal{T}) = \bsym{\mathcal{R}}_k^T \bsym{\mathcal{Q}}^T \bsym{\mathcal{S}} \, \bsym{\mathcal{Q}} \, \bsym{\mathcal{R}}_l,
\end{align}
an overlap over the positive definite matrix $\bsym{\mathcal{Q}}^T \bsym{\mathcal{S}} \, \bsym{\mathcal{Q}}$. In block-form, we moreover have
\begin{align}
\bsym{\mathcal{S}} = \begin{pmatrix} 1 & 0 \\ 0 & \b{S} \end{pmatrix}.
\end{align}

To derive a useful expression for $f$, we separate the reference and excited contributions. Note that the left and right excited states satisfy
\begin{align}
\bsym{\mathcal{A}} \, \bsym{\mathcal{R}}_n = \omega_n \, \bsym{\mathcal{R}}_n, \quad \bsym{\mathcal{L}}_n^T \bsym{\mathcal{A}} = \omega_n \, \bsym{\mathcal{L}}_n^T,
\end{align}
where the the left ground state vector is determined from the multiplier equation\citep{Koch1990}
\begin{align}
\bsym{\mathcal{L}}_0^T = (1 \; \b{\overline{t}}^T), \quad \b{\overline{t}}^T \J = - \veta^T.
\end{align}
From this it is straight-forward to show that
\begin{align}
\bsym{\mathcal{L}}_n^T = (0 \; \b{l}_n^T), \quad \bsym{\mathcal{R}}_n = \begin{pmatrix} - \b{\overline{t}}^T \b{r}_n \\ \b{r}_n \end{pmatrix}, \quad \J \, \b{r}_n = \omega_n \, \b{r}_n, \quad \b{l}_n^T \, \J = \omega_n \, \b{l}_n^T, \quad n > 0.
\end{align}
Substituting the block forms of $\bsym{\mathcal{Q}}$ and $\bsym{\mathcal{R}}_n$ into the expression for $f$ gives
\begin{align}
f(\mathcal{T}) = \b{r}_k^T ( \mathcal{N} \, \b{\overline{t}} \, \b{\overline{t}}^T + \b{Q}^T \b{S} \; \b{Q} - \b{Q}^T \b{S} \; \b{q} \, \b{\overline{t}}^T - \b{\overline{t}} \, \b{q}^T \b{S} \; \b{Q} ) \, \b{r}_l = 0, \label{eq:overlap}
\end{align}
where $\mathcal{N} = 1 + \vq^T \b{S} \; \vq$. This is the implemented expression for the generalized overlap. The vector $\b{q}$ can be evaluated as
\begin{align}
q_{ai} = t_i^a, \quad q_{aibj} = \frac{1}{1 + \delta_{ai,bj}} (t_{ij}^{ab} + t_i^a t_j^b).
\end{align}
The transformation $\b{y} = \b{Q} \, \b{x}$ is
\begin{align}
y_{ai} = x_{ai}, \quad y_{aibj} = x_{aibj} + \frac{1}{1+\delta_{ai,bj}}(t_j^b \, x_{ai} + t_i^a \, x_{bj}).
\end{align}
The transformation $\b{y} = \b{Q}^T \, \b{x}$ can be written
\begin{align}
y_{ai} = x_{ai} + \sum_{ck} t_k^c \, x_{aick}, \quad y_{aibj} = x_{aibj}.
\end{align}
Finally, $\b{y} = \b{S} \, \b{x}$ can be written
\begin{align}
y_{ai} = 2 \, x_{ai}, \quad y_{aibj} = 2 \, (1 + \delta_{ai,bj}) (2 \, x_{aibj} - x_{ajbi}).
\end{align}

\begin{table}[htb]
\caption{\footnotesize Energies obtained by various triple excitations $\tau_{IJK}^{ABC}$ for the geometry $R_{\ce{OH}} = 1.14$ Å, $R_{\ce{OF}} = 1.32$ Å, and ${\theta_{\ce{HOF}} = 91.0^\circ}$. We list CC2, SCCSD, CCSD, and CC3 energies given in Hartrees and the aug-cc-pVDZ basis. Dashes (--) denote that we were unable to converge the SCCSD equations. The first excitation listed in the table was used in the paper.}
\begin{tabular}{ccccccc}
\toprule
$\begin{pmatrix}A & B & C \\ I & J & K\end{pmatrix}$ & $E_0$ & $E_1$ & $E_2$ & $\omega_1$ & $\omega_2$ & $\zeta$ \\
\midrule
$\begin{pmatrix}10 & 2 & 2 \\ 7 & 5 & 8 \end{pmatrix}$ & $-175.1605$ & $-174.8452$ & $-174.8440$ & $0.3153$ & $0.3165$ & $1.6688$ \\ 
\midrule
$\begin{pmatrix}10 & 2 & 8 \\ 7 & 5 & 8 \end{pmatrix}$ & $-175.1619$ & $-174.8445$ & $-174.8435$ & $0.3174$ & $0.3184$ & $-0.6551$ \\ 
\midrule
$\begin{pmatrix}10 & 2 & 10 \\ 7 & 5 & 8 \end{pmatrix}$ & $-175.1611$ & $-174.8467$ & $-174.8431$ & $0.3144$ & $0.3180$ & $2.2880$ \\ 
\midrule
$\begin{pmatrix}8 & 2 & 2 \\ 7 & 5 & 8 \end{pmatrix}$ & $-175.1605$ & $-174.8448$ & $-174.8434$ & $0.3157$ & $0.3170$ & $1.6531$ \\ 
\midrule
$\begin{pmatrix}8 & 2 & 8 \\ 7 & 5 & 8 \end{pmatrix}$ & -- & -- & -- & -- & -- & -- \\ 
\midrule
$\begin{pmatrix}8 & 2 & 10 \\ 7 & 5 & 8 \end{pmatrix}$ & $-175.1613$ & $-174.8448$ & $-174.8436$ & $0.3165$ & $0.3176$ & $0.4853$ \\ 
\midrule
$\begin{pmatrix} 3 & 1 & 1 \\ 8 & 8 & 5 \end{pmatrix}$ & $-175.1623$ & $-174.8455$ & $-174.8430$ & $0.3168$ & $0.3193$ & $2.7638$ \\ 
\midrule
$\begin{pmatrix} 10 & 1 & 1 \\ 7 & 5 & 8 \end{pmatrix}$ & $-175.1639$ & $-174.8445$ & $-174.8416$ & $0.3195$ & $0.3223$ & $-1.3795$ \\ 
\midrule
$\begin{pmatrix} 10 & 1 & 2 \\ 7 & 5 & 8 \end{pmatrix}$ & $-175.1616$ & $-174.8451$ & $-174.8441$ & $0.3165$ & $0.3175$ & $-0.4178$ \\ 
\midrule
$\begin{pmatrix} 10 & 1 & 3 \\ 7 & 5 & 8 \end{pmatrix}$ & $-175.1639$ & $-174.8438$ & $-174.8428$ & $0.3201$ & $0.3211$ & $-1.0914$ \\ 
\midrule
$\begin{pmatrix} 8 & 1 & 1 \\ 7 & 5 & 8 \end{pmatrix}$ & $-175.1597$ & $-174.8469$ & $-174.8453$ & $0.3127$ & $0.3144$ & $ 1.7677$ \\ 
\midrule
$\begin{pmatrix} 8 & 1 & 2 \\ 7 & 5 & 8 \end{pmatrix}$ & -- & -- & -- & -- & -- & --  \\ 
\midrule
CC2 & $-175.1590$ & $-174.8600$ & $-174.8440$ & $0.2990$ & $0.3150$ & -- \\
CCSD & $-175.1619$ & $-174.8451$ & $-174.8437$ & $0.3168$ & $0.3181$ & -- \\
CC3 & $-175.1745$ & $-174.8585$ & $-174.8558$ & $0.3160$ & $0.3187$ & -- \\
\bottomrule
\end{tabular} \label{tab:hof_tests}
\end{table}

\subsubsection*{Implementation tests} The following tests of were performed. 
\begin{itemize}
\item For a given value of $\zeta$, the excitation energies $\omega_n$ derived from the left and right eigenvalue problems are identical. Thus, $\J$ and $\J^T$ are internally consistent.
\item For a given value of $\zeta$, the identity\citep{Koch1990}
\begin{align}
A_{\mu\nu} = \frac{\partial \Omega_\mu}{\partial t_\nu},
\end{align}
evaluated by numerical differentiation and by transformation of elementary basis vectors, is satisfied for the \ce{LiH} molecule. Thus, $\bsym{\Omega}$ and $\J$ are internally consistent. Moreover, by the previous test, $\bsym{\Omega}$, $\J$, and $\J^T$ are internally consistent.
\item We confirmed that
\begin{align}
\b{\overline{t}}^T = \frac{\vq^T \b{S} \, \b{Q}}{1 + \vq^T\,\b{S} \, \vq},
\end{align}
is satisfied for \ce{H2}, indicating that the implementation of $\vq$ and $\b{Q}$ are correct. This identity can be shown to be valid from the completeness of $T = T_1 + T_2$.
\item For two states of the same symmetry in \ce{H2}, we confirmed that 
\begin{align}
\bsym{\mathcal{R}}_k^T \bsym{\mathcal{Q}}^T \, \bsym{\mathcal{S}} \, \bsym{\mathcal{Q}} \, \bsym{\mathcal{R}}_l = 0
\end{align}
is correct to the accuracy that the amplitudes and eigenvectors are converged.
\item Noting that 
\begin{align}
Q_{\mu\nu} = Q_{\mu\nu}(\mathcal{T}) = \Tbraket{\Phi_\mu}{e^\mathcal{T}}{\Phi_\nu}, \quad Q_{\mu\nu}^{-1} = \Tbraket{\Phi_\mu}{e^{-\mathcal{T}}}{\Phi_\nu} = Q_{\mu\nu}(-\mathcal{T}),
\end{align}
we confirmed that $\b{Q}(\mathcal{T}) \, \b{Q}(-\mathcal{T})$ and $\b{Q}^T(\mathcal{T}) \, \b{Q}^T(-\mathcal{T})$ equal the identity matrix $\b{I}$ for the implemented $\b{Q}$ and $\b{Q}^T$ transformations.
\item We confirmed that $\vq^T \b{S} \, \b{Q} \, \b{x}$ and $\b{x}^T \b{Q}^T \b{S} \, \vq$ are equal, indicating that $\b{Q}$ and $\b{Q}^T$ are internally consistent.
\end{itemize}

\subsubsection*{The algorithm} We adopt a self-consistent approach. For a fixed triple amplitude, $t_{\mu_3}$, the ground state amplitude equations, $\bsym{\Omega} = 0$, are first solved for $t_{\mu_1}$ and $t_{\mu_2}$. Given the converged singles and doubles amplitudes, the excited states $\b{r}_i$ are found and the overlap $f(\mathcal{T})$ evaluated. A Newton-Raphson algorithm, designed to locate a zero of the overlap function $f(\mathcal{T})$, then provides the next triple amplitude $t_{\mu_3}$. These steps are repeated until both the ground state equations are satisfied and $f(\mathcal{T}) = 0$. See Algorithm 1.

\subsection*{The cluster operator}
In the similarity constrained theory, a particular triple excitation is used. In Table \ref{tab:hof_tests}, we list the energies obtained for hypofluorous acid using twelve different triple excitations.

\subsection*{Left generalized orthogonality} We have chosen to enforce orthogonality of the right eigenvectors in the present study, but, for completeness, we list the equations necessary to enforce orthogonality among the left eigenvectors here. The left overlap $f_L$ can be written
\begin{align}
f_L(\mathcal{T}) = \bsym{\mathcal{L}}_k^T (\bsym{\mathcal{Q}}^T \bsym{\mathcal{S}} \, \bsym{\mathcal{Q}})^{-1}  \bsym{\mathcal{L}}_l.
\end{align}  
By block inversion of $\bsym{\mathcal{Q}}$ and $\bsym{\mathcal{S}}$, we have
\begin{align}
\bsym{\mathcal{Q}}^{-1} = \begin{pmatrix} 1 & 0 \\ - \b{Q}^{-1} \vq & \b{Q}^{-1} \end{pmatrix}, \quad \bsym{\mathcal{S}}^{-1} = \begin{pmatrix} 1 & 0 \\ 0 & \b{S}^{-1} \end{pmatrix}.
\end{align}
and therefore
\begin{align}
\begin{split}
(\bsym{\mathcal{Q}}^T \bsym{\mathcal{S}} \, \bsym{\mathcal{Q}})^{-1} &= \begin{pmatrix} 1 & 0 \\ - \b{Q}^{-1} \vq & \b{Q}^{-1} \end{pmatrix} \begin{pmatrix} 1 & 0 \\ 0 & \b{S}^{-1} \end{pmatrix} \begin{pmatrix} 1 & -\vq^T \b{Q}^{-T} \\ 0 & \b{Q}^{-T} \end{pmatrix} \\
&= \begin{pmatrix} 1 &  -\vq^T \b{Q}^{-T} \\ - \b{Q}^{-1} \vq & \b{Q}^{-1} (\b{S}^{-1} + \vq \, \vq^T) \b{Q}^{-T} \end{pmatrix}.
\end{split}
\end{align}
Now, because the reference term of $\bsym{\mathcal{L}}_k$ is zero (it is orthogonal to $\bsym{\mathcal{R}}_0$), we can write
\begin{align}
f_L(\mathcal{T}) = \b{l}_k^T \, \b{Q}^{-1} (\b{S}^{-1} + \vq \, \vq^T) \b{Q}^{-T} \, \b{l}_l.
\end{align}

\subsection*{Hypoflorous acid: intersection point, normal modes, seam, and branching plane vectors}
\subsubsection*{Intersection point}
The studied intersection geometry hypofluorous acid is
\begin{align}
R_{\ce{OH}} = 1.1400000 \; \text{\AA}, \quad R_{\ce{OF}} = 1.3184215 \; \text{\AA}, \quad \theta_{\ce{HOF}} = 91.0585000^\circ,
\end{align}
where the triple amplitude is $\zeta = 1.6178960762$. For this geometry and amplitude, the overlap is below $10^{-6}$ and the energies degenerate to $10^{-6}$:
\begin{align}
    \omega_1 = 0.3163264850 \; \text{Hartrees}, \quad \omega_2 = 0.3163274291 \; \text{Hartrees}.
\end{align}

\subsubsection*{Normal modes} We specify the nuclear cartesian coordinates as
\begin{align}
\nucl = (O_x,O_y,O_z,H_x,H_y,H_z,F_x,F_y,F_z)
\end{align}
in the following. From a vibrational Hartree-Fock calculation using the Cfour program\citep{cfour}, we obtained a set of normal modes,
\begin{align}
   \b{Q}_1 &= (0.7370,-0.0071,0.0000,
          -0.0070,0.0211,0.0000,
          -0.6750,0.0017,0.0000), \\
   \b{Q}_2 &= (0.0749,-0.2371,0.0000,
          -0.7773,0.5604,0.0000,
           0.1103,0.0885,0.0000), \\
   \b{Q}_3 &= (0.0773,0.1431,0.0000,
          -0.5890,-0.7874,0.0000,
          0.0647,0.0500,0.0000).
\end{align}
The cartesian coordinate vector at the intersection point is
\begin{align}
\begin{split}
\nucl_0 = (& -1.308090861096777,0.135129007453069,0.000000000000000, \\
            & -1.470679327231972,-2.013015024947860,0.000000000000000, \\
            &1.179309062794356,-0.006980060310293,-0.000000000000000).
\end{split}
\end{align}

\subsubsection*{The seam and branching plane vectors}
By performing small displacements in the normal modes $\b{Q}_i$, we found the basis $\{\b{g},\b{h}\}$ of the branching plane and the seam $\b{s}$. In the basis of the normal modes, 
\begin{align}
\b{g} &= (0.972636, -0.160489, 0.167996),\\
\b{s} &= (0.220983, 0.407941, -0.885862),\\
\b{h} &= (0.073639, 0.898745, 0.432243).
\end{align}
Note that these vectors are expected to have some numerical imprecision, arising from the finite number of fixed-point calculations on which they are based ($\b{s}$ and $\b{g}$ are nearly but not perfectly orthogonal, $89.96^\circ$) and that the $\b{Q}_i$ is given to four decimal places.

\end{document}